# ConFiG: Contextual Fibre Growth to generate realistic axonal packing for diffusion MRI simulation

Ross Callaghan[†], Daniel C. Alexander, Marco Palombo*, Hui Zhang*

Centre for Medical Image Computing and Department of Computer Science, University College London, London, UK

[†]Corresponding author: ross.callaghan.16@ucl.ac.uk; UCL Centre for Medical Image Computing, 90 High Holborn, London, WC1V 6LJ, UK

*Joint senior author

**Abstract**

This paper presents Contextual Fibre Growth (ConFiG), an approach to generate white matter numerical phantoms by mimicking natural fibre genesis. ConFiG grows fibres one-by-one, following simple rules motivated by real axonal guidance mechanisms. These simple rules enable ConFiG to generate phantoms with tuneable microstructural features by growing fibres while attempting to meet morphological targets such as user-specified density and orientation distribution. We compare ConFiG to the state-of-the-art approach based on packing fibres together by generating phantoms in a range of fibre configurations including crossing fibre bundles and orientation dispersion. Results demonstrate that ConFiG produces phantoms with up to 20% higher densities than the state-of-the-art, particularly in complex configurations with crossing fibres. We additionally show that the microstructural morphology of ConFiG phantoms is comparable to real tissue, producing diameter and orientation distributions close to electron microscopy estimates from real tissue as well as capturing complex fibre cross sections. Signals simulated from ConFiG phantoms match real diffusion MRI data well, showing that ConFiG phantoms can be used to generate realistic diffusion MRI data. This demonstrates the feasibility of ConFiG to generate realistic synthetic diffusion MRI data for developing and validating microstructure modelling approaches.

**Highlights**

- We present ConFiG, a biologically motivated numerical phantom generator for white matter
- ConFiG produces phantoms with state-of-the-art density and realistic microstructure
- Diffusion MRI simulations in ConFiG phantoms are comparable to real dMRI signals

**Abbreviations**

CAM: cell adhesion molecules, CC: corpus callosum, ConFiG: contextual fibre growth, dMRI: diffusion MRI, EM: electron microscopy, FOD: fibre orientation distribution, HCP: human connectome project, IC: internal capsule, TC: three crossing region, WM: white matter

## 1. Introduction

Numerical phantoms play a valuable role in the development and validation of many magnetic resonance imaging (MRI) techniques. In particular, numerical phantoms are often used when developing diffusion MRI (dMRI) microstructure imaging techniques where simulations of the dMRI signal in phantoms with known microstructural properties are used in lieu of an *in vivo* ground truth measure of microstructure. For instance, many studies employing dMRI modelling to study microstructural features of white matter (WM) use numerical phantoms as part of the validation process, comparing parameter estimates from fitting their models to the known ground truth from the phantoms e.g. (Jelescu and Budde, 2017; Li et al., 2019; Nilsson et al., 2017, 2010; Scherrer et al., 2016; Tariq et al., 2016; Xu et al., 2014; Zhang et al., 2012). Some recent works directly estimate microstructural features using fingerprinting techniques and machine learning to match simulated signals and the corresponding ground truth microstructure of the numerical phantom to the measured signal (Hill et al., 2019; Nedjati-Gilani et al., 2017; Palombo et al., 2018a; Rensonnet et al., 2018). As well as affecting the dMRI signal, axonal configuration also influences other MR techniques such as susceptibility-weighted imaging (Lee et al., 2010; Li et al., 2012). For instance, Xu et al., (2018) recently used simulations to show that using realistic axonal models rather than simple circular cylinders affects the MR signal. Therefore, it is important to the MRI community to generate realistic WM numerical phantoms which accurately capture microstructural features in order to get realistic simulated signal.

Typically, however, there is a mismatch between the complexity of true brain tissue microstructure and the models used in simulation, with simulations simplifying the microstructure. One one hand, *ex vivo* electron microscopy (EM) studies have revealed the high complexity of real axonal morphology (Abdollahzadeh et al., 2019; Lee et al., 2019; Salo et al., 2018). Reconstructions of axons from these studies show that real WM contains axons with complex morphologies on an individual axon basis such as undulation, beading and non-circular cross sections, as well as non-trivial configurations including orientation dispersion and crossing bundles. On the other hand, the models used in simulation studies often represent axons in WM using simplistic geometrical representations such as parallel cylinders with uniform (Fieremans et al., 2010; Ford and Hackney, 1997; Nilsson et al., 2010, 2009) or polydisperse (Alexander et al., 2010; Hall and Alexander, 2009) radii. Some studies investigate the effect of differing configurations of fibres such as simple crossing (Rensonnet et al., 2017) and planar dispersed (Zhang et al., 2011) fibre bundles. A few groups generate WM numerical phantoms with complex fibre configurations for the application to tractography (Close et al., 2009; Neher et al., 2014); however realistic microstructural morphology is not the focus of these approaches. Other studies introduce more microstructural complexity into the numerical phantoms, typically only considering one mode of morphological variation at a time; some examples of this include harmonic beading (Budde and Frank, 2010; Landman et al., 2010), spines (Palombo et al., 2018b), undulation (Brabec et al., 2019; Nilsson et al., 2012) and myelination (Brusini et al., 2019).

Recently, a number of groups have attempted the challenge of combining these features to generate phantoms approaching the morphological complexity and density of real tissue. The most common approach to this is the packing of fibres into densely packed configurations (Close et al., 2009; Ginsburger et al., 2019, 2018; Rafael-Patino et al., 2018). The typical approach, as taken in the state-of-the-art MEDUSA algorithm (Ginsburger et al., 2019), is to generate a set of overlapping fibres decomposed into small segments and iteratively refine their positions to remove the overlap between them. Despite their recent progress, further advance of this class of techniques may be limited, because nature does not create fibres before attempting to pack them together. Instead, real axons are guided by chemical cues and fit into available space as they grow (Lowery and Vactor, 2009; Price et al., 2017). Mimicking the natural fibre genesis may prove important for building more realistic phantoms.

To this end, we propose Contextual Fibre Growth (ConFiG), an approach to generate WM numerical phantoms that emulates natural fibre growth. ConFiG generates WM numerical phantoms by growing fibres one-by-one, mimicking a set of key mechanisms which govern real axonal growth. A preliminary implementation of ConFiG was presented in (Callaghan et al., 2019). We assess the performance of ConFiG by measuring the impact of each of the biologically inspired mechanisms on the achievable phantom density and comparing against state-of-the-art MEDUSA phantoms. To test how realistic ConFiG phantoms are, we compare the microstructural properties of the phantoms to measured data from electron microscopy and compare simulated dMRI signal in the phantoms to real dMRI data.

The rest of the paper is organized as follows: Section 2 describes the ConFiG algorithm, Section 3 details the experiments outlined above and Sections 4 and 5 summarise the contributions and discuss future work.

## 2. Methods

In this section we describe the ConFiG algorithm, beginning with an overview of the main components in the growth algorithm. We then describe the biological mechanisms motivating ConFiG and how each of these are implemented to give the final ConFiG algorithm.

### 2.1. Overview of the ConFiG algorithm

Given a set of morphological input parameters (target density, orientation distribution and diameter distribution), ConFiG generates a densely packed set of fibres by growing each fibre following a set of biologically motivated rules. The generation of ConFiG numerical phantoms happens in three main steps:

- STEP 1: Generate initial growth configuration from user inputs
- STEP 2: Grow the fibres using ConFiG growth algorithm
- STEP 3: Generate 3D meshes for dMRI simulation

Each of these steps are discussed in detail below.

First, Step 1 is broken down in to three substeps as outlined in Figure 1:

STEP 1.1: Generate fibre starting points (Figure 1a-b). To generate a starting point for each fibre to grow from, ConFiG packs circles with the desired diameter distribution up to the target density in 2D, following the approach taken in (Hall and Alexander, 2009).

STEP 1.2: Generate fibre target points (Figure 1c). To encode the desired orientation distribution, each fibre has a direction drawn from the target distribution which gives a target point for the fibre to grow towards.

STEP 1.3: Generate growth network nodes (Figure 1d). ConFiG uses a network of pseudorandomly placed nodes to sample the space and encode which regions are occupied by existing fibres. This simplifies collision checking making growth more efficient than a direct collision detection approach involving growing each fibre one small step at a time and checking collisions with existing fibres (Callaghan et al., 2019).

Second, Step 2, the main growth algorithm, is broken down into a series of substeps as outlined in Figure 2:

STEP 2.1: Create growth network (Figure 2a&b). In order to encode which nodes a fibre can move to from any other node, the growth nodes are connected using the Delaunay triangulation.

STEP 2.2: Grow one fibre step (Figure 2 c-e). Fibres grow one-by-one in a random order along this network towards their target points while avoiding existing fibres. During growth, a fibre must choose in which direction it should grow. This direction is chosen in ConFiG by following a cost function motivated by biological axonal guidance mechanisms (Figure 2d), described Sections 2.3 and 2.6.

STEP 2.3: Update the network (Figure 2 f). The growth network is updated in order to store the information about the space this fibre is occupying so that future fibres can avoid it. The simplest way to do this is to store the minimum distance from each node in the network to any existing fibre as in (Callaghan et al., 2019). Additionally, another biologically motivated network updating strategy is described in Section 2.5.

STEP 2.4: Repeat steps 4&5 until fibre reaches target (Figure 2g). By default in ConFiG, each fibre will grow completely before the next one starts, meaning that step 5 only needs to be performed once the fibre has finished growing. If fibres are allowed to grow concurrently, step 5 must be performed after each growth step.

STEP 2.5: Repeat steps 4-6 for remaining fibres (Figure 2h-i). As noted in Figure 2 (e-h), as the network is updated, more and more nodes become inaccessible making the network sparser. This means that some fibres may reach a point from which they cannot grow any further and will become stuck. Biologically inspired mechanisms designed to address this point are described in Sections 2.4 and 2.5.

Finally, Step 3, the meshing procedure, is briefly described below and in further detail in Section 2.8:

STEP 3: Generate 3D fibre meshes. After the growth process, each fibre will be represented by a series of connected 3D points and corresponding diameters at each point. In order to simulate diffusion MRI signals, these fibre skeleta need to be turned into 3D meshes. ConFiG uses a meshing procedure designed to eliminate overlap between fibres.

The basic ConFiG growth algorithm described here is illustrated in Figure 2, with an animation of the algorithm in Supplementary Video 1. The remainder of this section outlines the biological process governing real axonal growth, and how these processes motivated the final implementation of the ConFiG algorithm.

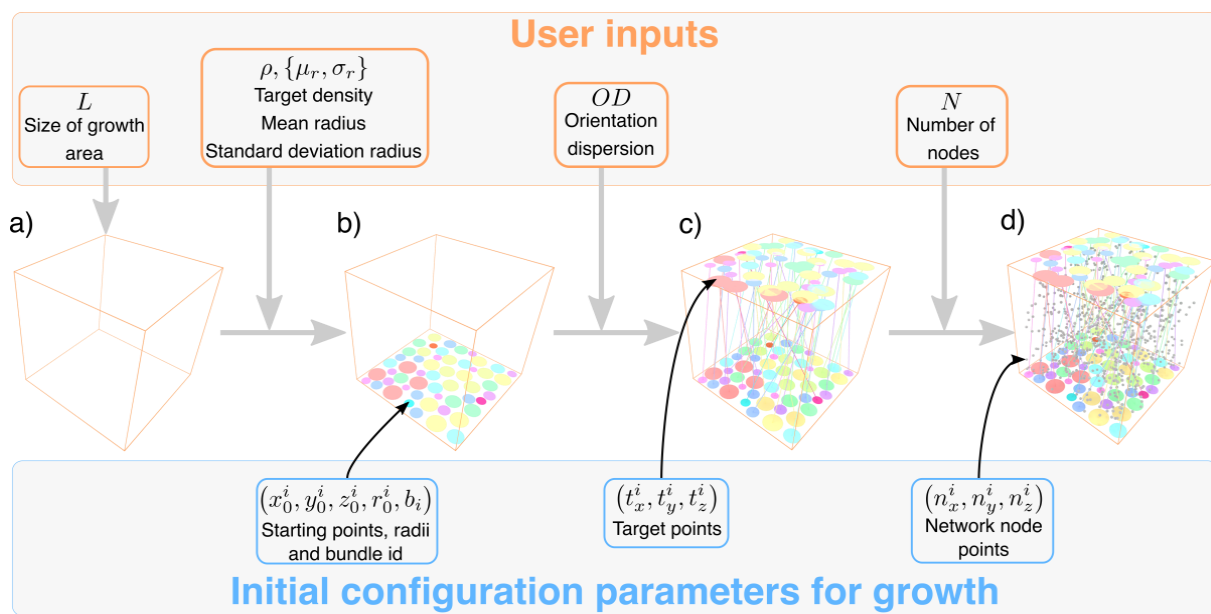

*Figure 1 Inputs to the ConFiG algorithm for the single bundle case. L defines the size of the area of that the growth will take place in. The target density and fibre radius distribution govern the generation of starting points for each fibre by packing in 2D. Orientation dispersion parameters govern the generation of target points corresponding to each starting point. N*

*defines the number of nodes to use when generating the network. In the case of multiple bundles, starting and target points are generated for each bundle and then combined into the same space which is filled with nodes for the network.*

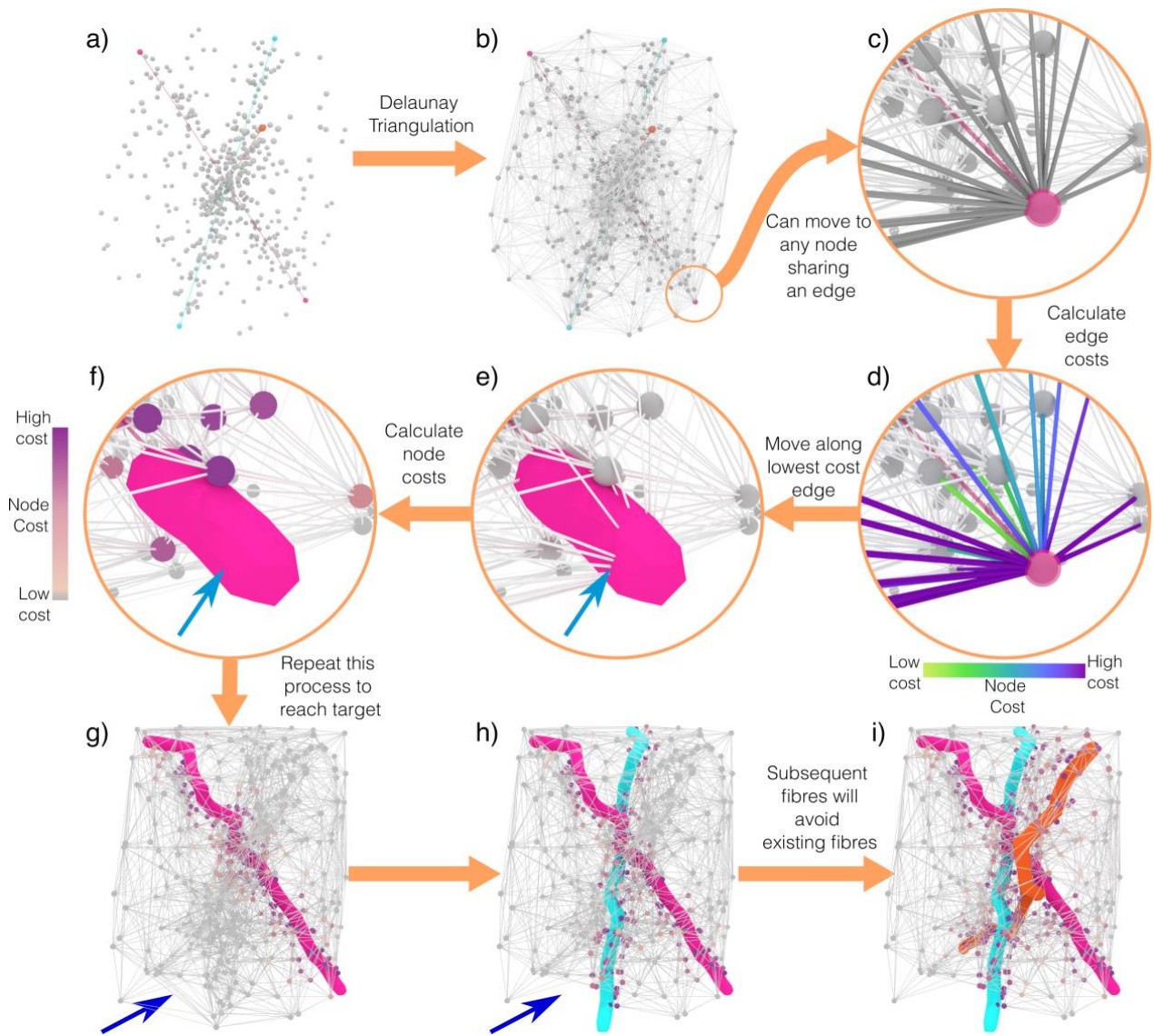

*Figure 2 Overview of the basic growth algorithm in ConFiG. In this example, three fibres are shown with a growth network that only contains relevant nodes for the sake of visualisation. From the set of nodes, a network is constructed using the Delaunay triangulation. Each fibre then grows from node to node, along any edge connected to the current node. The node moved to will be the node with the lowest cost. Once a fibre segment has grown, the network nodes are updated to store information about which nodes are occupied or near to an existing fibres. This contributes to the cost function for any future fibres, penalising moving to nodes too close to existing fibres. It is not possible to move to any node now inside a fibre as indicated by the removal of this edges from the network (pairs of blue arrows show where this is happening). The next fibres grow, now avoiding existing fibres until all fibres have finished. See Supplementary Video 1 for an animation of this algorithm.*

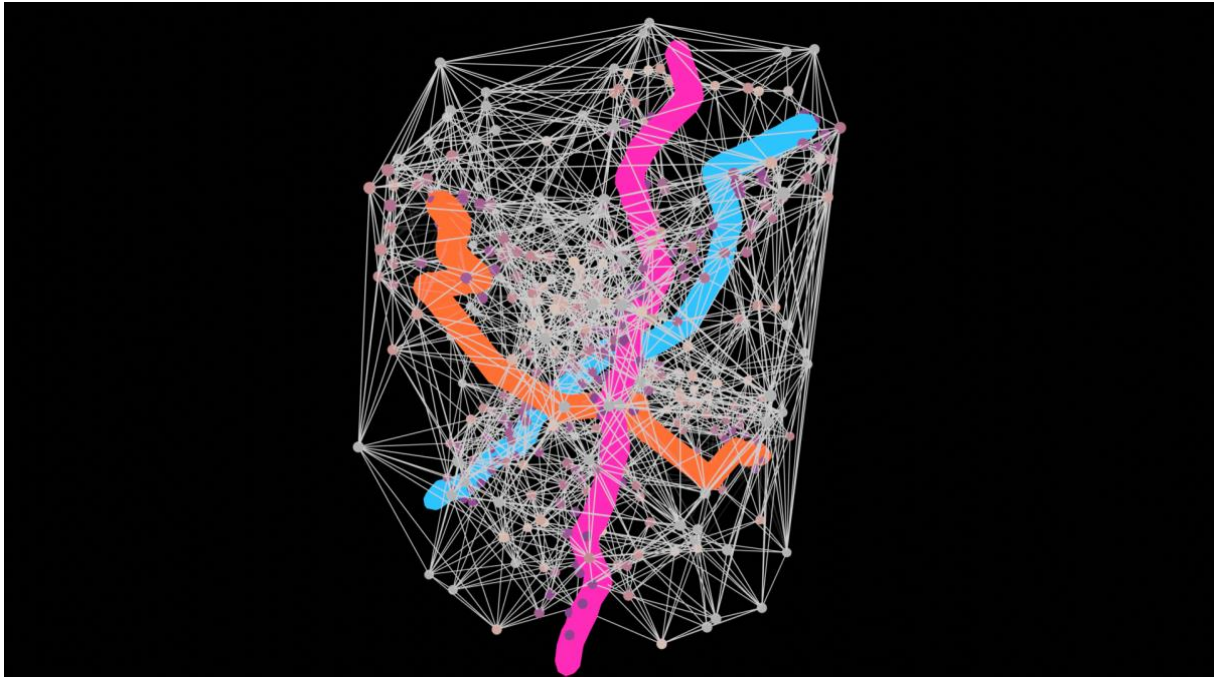

*Video 1 Illustration of the ConFiG growth algorithm overviewed in Section 2.1 and Figures 1 and 2. Starting and target points for each fibre are generated before the growth network is created and the fibres grow along it following a biologically motivated cost function. (Video not available on arxiv, contact corresponding author).*

**2.2. Biological motivation for ConFiG**

Axons grow following chemical cues in their environment through various mechanisms which either attract or repel fibres to guide their growth (Dent et al., 2011; Lowery and Vactor, 2009; Mortimer et al., 2008; Polleux and Snider, 2010; Price et al., 2017; Rauch et al., 2013; Sakisaka and Takai, 2005). In an attempt to emulate real axonal growth, mechanisms motivated by the following guidance processes have been integrated into ConFiG:

- Chemoattraction – the process by which fibres are attracted to diffusible chemical cues in their environment (Mortimer et al., 2008; Price et al., 2017).
- Fibre collapse – a response to a chemorepulsive source whereby a fibre withdraws and regrows in a different direction (Rauch et al., 2013).
- Cell adhesion molecules – chemical signals on the surface of cells which guide axons that come into contact with them (Sakisaka and Takai, 2005).
- Fasciculation – the process by which multiple axons come together to form bundles (Price et al., 2017; Šmít et al., 2017).

The following sections detail how mechanisms motivated by these biological processes are implemented in ConFiG while Figures 3-5 illustrate these biological processes alongside their ConFiG counterparts.

## 2.3. Chemoattraction

As mentioned in Section 2.1, as a fibre grows it must choose in which direction it will move. One of the main processes governing the guidance of real axons is chemotropism; a process by which axons respond to diffusible chemical cues in their environment. One key chemotropic mechanism is chemoattraction, in which fibres are attracted along a chemical gradient towards a target region (Price et al., 2017).

To approximate this chemoattractive mechanism, each fibre is encouraged to grow towards its target point (i.e. the target point acts like a chemoattractive source). From any node in the growth network, the fibre will move along an edge that takes it towards its target while avoiding existing fibres according to a cost function (Callaghan et al., 2019). The chemoattractive mechanism and its ConFiG counterpart are illustrated in Figure 3a.

From a starting node, $s$, the candidate nodes, $c$, that the fibre can move to are any nodes that share an edge with $s$. In addition to its position, each network node stores the maximum diameter, $d_c$, that can be sustained at that node without intersecting another fibre. The fibre will move to a candidate node according to a cost function consisting of two terms; $l_t$, which penalises moving away from the target point, $t$, and $l_d$, which penalises moving to a position where $d_c$ is low meaning that the fibre will have to shrink. The cost function for a fibre at a position, $s$, to move to a candidate node, $c$, given a target point, $t$, is (Callaghan et al., 2019)

$$l = l_t + f l_d , \qquad (1)$$

where

$$l_t = \frac{1}{2} \cdot \frac{\|s - c\|}{1 + \|s - c\|} \cdot \left(1 - \frac{(c - s) \cdot (t - s)}{\|c - s\| \|t - s\|}\right), \qquad (2)$$

$$l_d = \max\left(0, \frac{1}{d_0}(d_0 - d_c)\right). \qquad (3)$$

Here, $d_0$ is the target diameter of the fibre and $f$ is a weighting factor between the two terms. In this work, $f$ is fixed to 0.2 to more strongly weight growth towards the target.

The next node for a fibre will be the candidate node which has the lowest cost according to Equation (1). This method of finding a path through the triangulation by choosing the lowest cost node at each position amounts to a greedy best-first pathfinding approach with a heuristic given by Equation (1).

Growing fibres along the network using just this chemoattractive mechanism is the minimal implementation of ConFiG that will generate substrates to try and meet the morphological inputs. There are some limitations to this minimal approach however; the greedy growth and the sparse sampling of the space means that fibres can grow into regions from which they cannot grow further and become stuck. Additionally, in this approach, fibres grow independently of one another, whereas real fibres grow forming bundles in the process known as fasciculation.

Sections 2.4-2.7 describe further mechanisms which were added to enable ConFiG to address these limitations in order to meet more complex morphological priors (e.g. high density and orientation dispersion together).

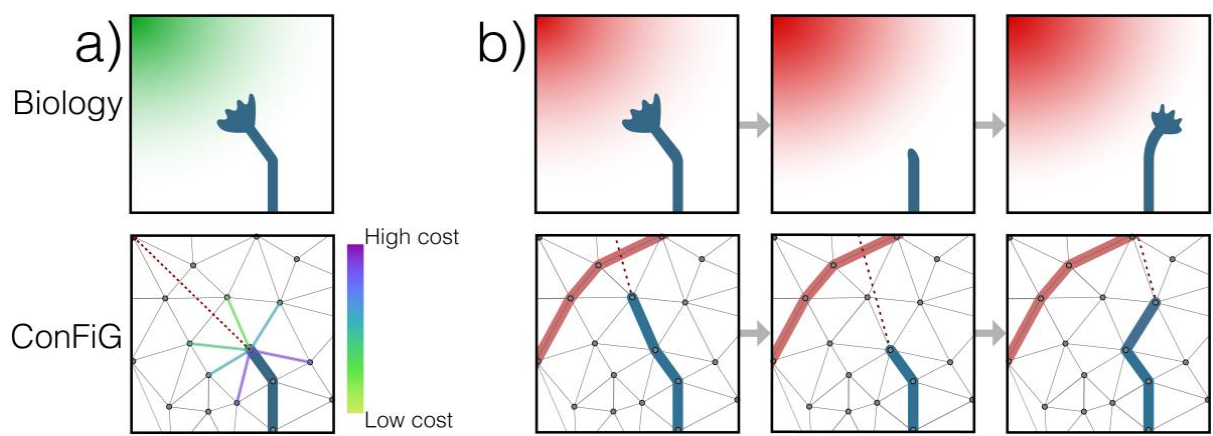

*Figure 3 Illustration of two of the biological motivations and how they are implemented in ConFiG. a) Growth towards the target is enforced by means of a cost function encouraging growth towards the target point. b) Fibre collapse is implemented by allowing the fibre to move backwards if it reaches a node from which there are no viable steps. The biological figures are adapted from (Price et al., 2017).*

## 2.4. Fibre collapse

As mentioned in Section 2.3, in ConFiG a fibre can become stuck when there are no possible next steps because all neighbouring nodes are inaccessible. In an attempt to ameliorate this a process mimicking fibre collapse was implemented, illustrated in Figure 3b.

In ConFiG fibre collapse, the fibre will move back by an initial distance, $d_0$, and regrow from there avoiding any nodes in the route it took previously. If the fibre becomes stuck again, it will move back by a further distance, $d_0 + \delta$, where $\delta$ is the additional distance to step back. This process is repeated until the fibre reaches the target or gets stuck a user-defined maximum number of times. In this work, $d_0 = 2$ μm and $\delta = 5$ μm in an approximation of the biological fibre collapse process investigated by Rauch et al. (Rauch et al., 2013) who show fibres collapsing up to 25μm back towards the soma. The maximum number of steps back is set to 5, meaning that the maximum step back is 27μm, in line with real fibres. If there is no possible route after 5 attempts then the fibre will stop growing and will be removed from the phantom. This process of removing stuck fibres means that the resulting substrate may not always have the same density as the input desired fibre density.

## 2.5. Dynamic Growth Network

In the preliminary implementation of ConFiG (Callaghan et al., 2019), the network nodes were initialised pseudorandomly within the growth region and once initialised, the growth network was static, meaning that the nodes and edges of the network were fixed. This limited the growth to the specific instantiation of the network and it could not adapt to where fibres were once they had grown. Furthermore, as illustrated in Figure 2, as fibres grow, many nodes become inaccessible due to being within fibres meaning that the network becomes gradually sparser.

A dynamic growth network was implemented to ameliorate these effects. Now, once a fibre has reached the target, a number of nodes, $N_{added}$, are generated around the path of the fibre. This gives a denser sampling of the space in regions in which fibres exist and serves to give subsequent fibres more nodes to use to grow along or around that fibre, helping to increase the achievable density by limiting the number of fibres which get stuck. In this work, where the dynamic network is used, $N_{added} = 2500$.

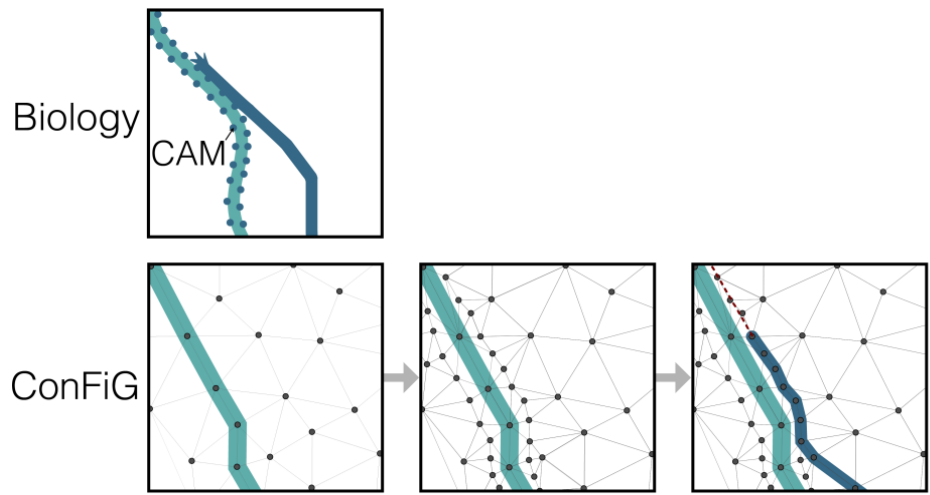

*Figure 4 Illustration of the contact guidance axonal growth mechanism and the dynamic growth network implemented in ConFiG. The dynamic growth network is implemented as a set of points added around each fibre after growth, enable future fibres to more easily grow along/around existing fibres.*

This is also loosely motivated by the contact guidance mechanism in which axons are attracted to or repelled by chemical cues on the surface of other cells, known as cell adhesion molecules (CAMs). Here, the added points act like CAMs meaning that a future fibre which grows can use these points near to the fibre to grow around or along it as if it were following contact guidance cues. Figure 4 shows how CAMs work in biological axonal growth alongside the ConFiG dynamic network, illustrating the parallels between the two.

## 2.6. Axon fasciculation

One particular role CAMs play is in axon fasciculation, the process in which axons follow a so-called pioneer axon closely, forming a bundle (Price et al., 2017; Sakisaka and Takai, 2005). To mimic the process of axon fasciculation, the term in the cost function penalising moving into regions in which the fibre had to shrink, $l_d$ (Equation (3)), was altered to be conditional on which fibre bundle is closest.

A fibre, $f$, with a target diameter, $d_0$, moving to a candidate node, $c$, which has a maximum sustainable diameter $d_c$ will now have $l_d$ given by:

$$l_d = \begin{cases} \max\left(0, \frac{1}{d_0}(d_0 - d_c)\right) & \text{if } b_c \neq b_f \\ \text{abs}\left(\frac{1}{d_0}(d_0 - d_c)\right) & \text{if } b_c = b_f \end{cases}, \qquad (4)$$

Where $b_f$ is an index identifying the bundle that fibre $f$ belongs to and $b_c$ is the index of the bundle that is closest to $c$ (i.e. the index of the bundle of the fibre that set $d_c$). This means that when $c$ is closest to the same bundle as $f$, the cost function penalises moving away from that bundle as well as shrinkage, whereas when the bundles differ, it only penalises shrinkage.

This new form of the cost function encourages fibres of the same bundle to stick together while still avoiding fibres of different bundles, inspired by the labelled pathway hypothesis, which states that axons join different fascicles based on different CAMs expressed on the fibres (Price et al., 2017). In this case, bundle indices $b_c$ and $b_f$ act like different identifying CAMs. Figure 5 shows how this fasciculation process is expected to happen in biology alongside how the improved cost function encourages a similar process in ConFiG.

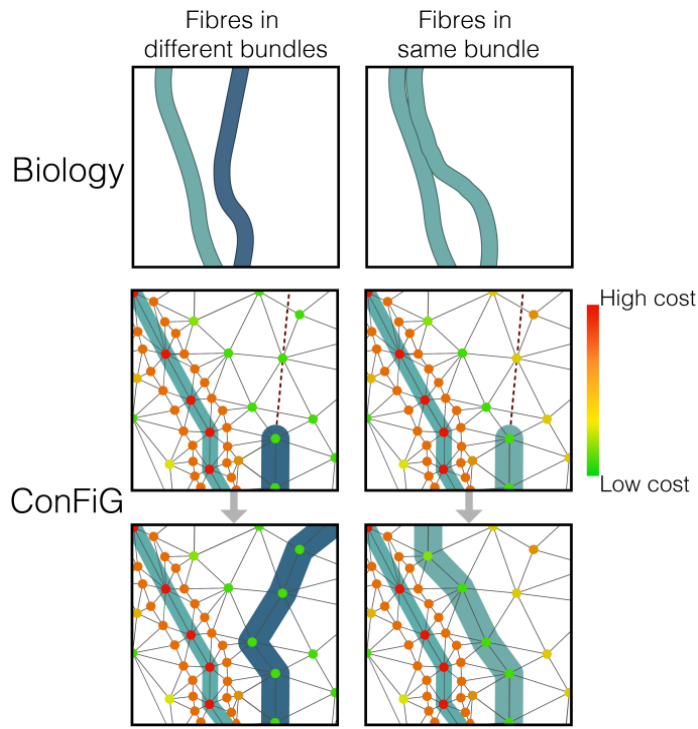

*Figure 5 Illustration of how the labelled pathway hypothesis is expected to work in biology and its ConFiG counterpart. Fasciculation is implemented using the cost function term in Eq. 1 which means that fibres in the same bundle are encouraged to stay close to one another.*

## 2.7. Global optimisation

Since the growth of fibres in ConFiG takes place on a discrete network of points, the final positions of fibre nodes may be suboptimal for achieving the maximum density. In other words, certain fibres' nodes may be closer to other fibres than they would ideally be in order to reach their target diameter (i.e. the fibre has had to shrink its diameter at that node).

To mitigate against this, a global optimisation step was added at the end of the growth in a procedure similar to MEDUSA (Ginsburger et al., 2019). For each point, $i$, that is part of a fibre, its nearest $n$ neighbours ($j \in NN(i)$) from other fibres are found; in this work $n = 10$. The distance to all of the neighbours is found and the point's position is updated from these distances according to the update vector, $\vec{u_i}$

$$\vec{u_i} = \sum_{j \in NN(i)} D(i,j) \cdot (\vec{p_i} - \vec{p_j}),$$

where $\vec{p_i}$ and $\vec{p_j}$ are the locations of point $i$ and $j$. $D(i,j)$ is the function determining whether the interaction is repulsive or attractive:

$$D(i,j) = \text{sgn}(r_i + r_j - \|\vec{p_i} - \vec{p_j}\|).$$

Here, sgn is the signum function and $r_i$ and $r_j$ are the target radii of point $i$ and $j$. The sum of these radii is the desired distance between the points since that means the fibres are just touching. $D(i,j)$ imposes that the force is repulsive if the points are closer together than the desired radius and attractive if they are further apart. The update vector is scaled such that if $\|\vec{u_i}\| > 0.2 r_i$, the update vector is rescaled so that $\|\vec{u_i}\| = 0.2 r_i$. This acts to prevent the update vector from becoming very large.

There is some biological evidence that this kind of interaction between fibres is important in the fasciculation process. The fasciculation process described in Section 2.4 relies on CAMs detected at the tip of a growing axon, however some studies provide evidence for fasciculation through interactions along axon shafts, known as zippering (Barry et al., 2010; Šmít et al., 2017; Voyiadjis et al., 2011). In zippering, nearby axon shafts attract one another to form more closely packed fascicles, which is a similar process to the global optimisation process in ConFiG.

**2.8. Creation of 3D meshes**

As mentioned in Section 2.1, following the growth process ConFiG fibres will be represented by a series of connected points and corresponding radii. To convert these skeleta into 3D meshes, the ConFiG meshing procedure uses Blender (https://blender.org) and is built on the SWC mesher addon (https://github.com/mcellteam/swc_mesher).

ConFiG meshes are constructed using Blender metaballs, an implicit surface representation which is the isosurface of a function; typically a function analogous to the electric potential from a point charge. When two metaballs come close to one another, the fields combine and the surfaces will merge to form a smooth surface. By placing a series of metaballs along the skeleton of each fibre, a smooth surface is formed for each fibre one-by-one as shown in Figure 6a.

When fibres are densely packed, the surfaces from neighbouring fibres may overlap. To account for this, a meshing procedure was developed in which fibres can deform around nearby fibres to avoid overlap. The metaball surface for one fibre is created as described above. This surface is then turned into a triangulated mesh, however the metaballs are retained. The metaball potential is then turned negative, meaning that rather than merging with any future nearby metaball surfaces, it will repel them, as shown in Figure 6b. This means that subsequent fibres which are meshed very close to, or overlapping with, existing fibres will deform organically to resolve the intersection, thus creating a series of completely non-intersecting fibre meshes which can be used by the dMRI simulator.

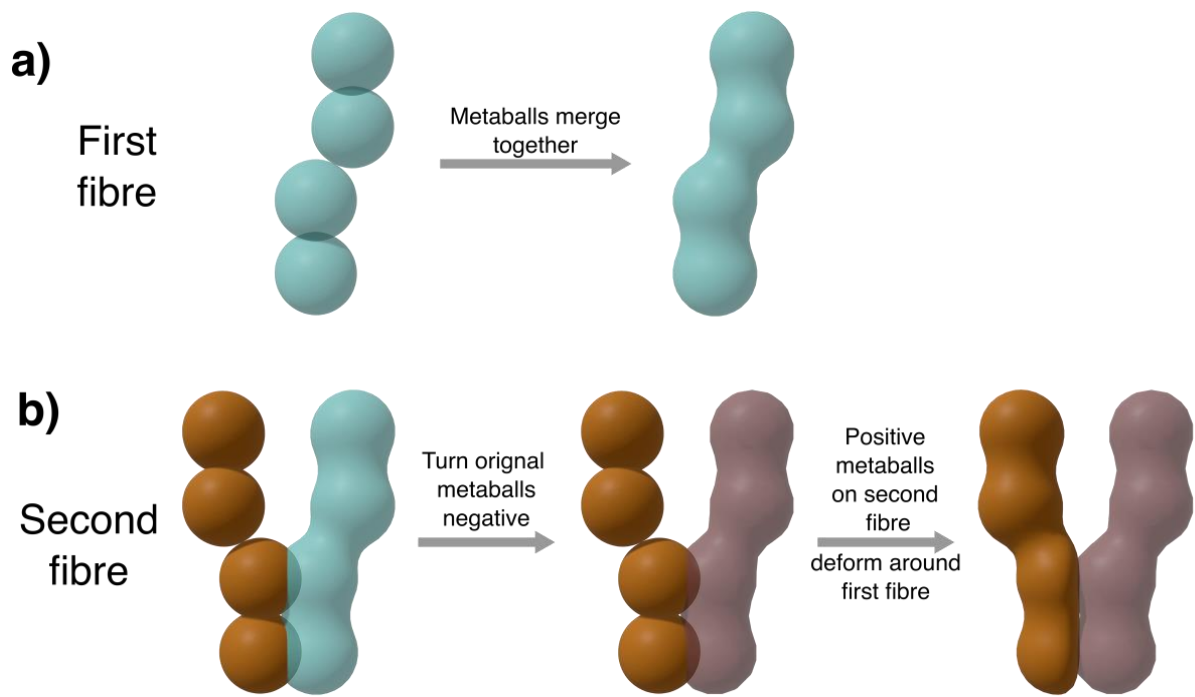

*Figure 6 Demonstration of the meshing procedure in ConFiG. The first fibre is created using metaballs to create a smooth surface. The second, and following fibres will be created using negative metaballs for any fibres that intersect in order to deform around them. Note that in practice, more spheres will be placed along the skeleton to create a smoother surface.*

## 3. Experiments

In order to assess the performance of ConFiG, a range of experiments were performed. The first set of experiments were performed in order to explore the impact of each of the biologically inspired growth mechanisms. Another set of experiments aimed to show that ConFiG is able to generate substrates with realistic microstructure by comparing generated substrates with real tissue. Additionally, the relationship between the user-specified target morphology and the final output morphology was investigated by comparing resulting phantoms to their inputs (target density and orientation distribution). Finally, a simulation experiment was performed to assess how well ConFiG phantoms can be used to generate realistic diffusion MRI data. The rest of this section outlines these experiments.

### 3.1 Testing the performance of ConFiG

In order to test how each of the biological mechanisms proposed in Section 2 impacted on the resulting phantoms, an experiment was devised to measure how phantoms changed when each mechanism was introduced. Four scenarios of interest were generated using several variants of the ConFiG algorithm that include these mechanisms either one at a time or all at once, attempting to grow phantoms as densely as possible:

- one bundle of parallel fibres, target density 75%
- one bundle with Watson distributed fibres ($\kappa$=8), target density 75%
- two perpendicular crossing bundles, intra-bundle target density 40%
- three mutually perpendicular crossing bundles, intra-bundle target density 30%

These target densities were chosen to ensure that the centre of the phantom (i.e. the crossing region for crossed bundles) had a high target density whilst ensuring that each bundle had a reasonable number of fibres to begin with (>50).

The ConFiG variants were tested by generating phantoms for each of the scenarios starting with the same initial conditions. Each phantom was generated 5 times with a different random seed and results averaged across the seeds.

To investigate the impact of the biological mechanisms on dMRI simulation, a comparison was made between real dMRI signals and simulations from ConFiG phantoms. The NODDI model (Zhang et al., 2012) was fitted to a WM ROI in the corpus callosum of a Human Connectome Project (HCP) (Van Essen et al., 2012) subject to provide sensible input parameters (target fibre density and orientation dispersion) for ConFiG to generate phantoms. We generate phantoms using the two extreme cases: the minimal growth case only using chemoattraction, and the complete ConFiG algorithm using all mechanisms. Whilst the random nature of ConFiG means that the resulting phantom will not have morphology exactly matching the input parameters, this approach ensured that the phantoms were reasonable for this proof of concept experiment.

The dMRI signal was simulated in the phantoms using Camino (Cook et al., 2006; Hall and Alexander, 2009) with identical simulation conditions in both cases and the measurement scheme corresponding to the HCP dMRI sequence(Stamatios N. Sotiropoulos et al., 2013).

## 3.2 Microstructural measures

In order to test how realistic the microstructure generated using ConFiG is, microstructural measurements of diameter distribution and orientation distribution were calculated using methods to be comparable with previous studies on ex-vivo tissue (Abdollahzadeh et al., 2019; Lee et al., 2019).

A centre line is generated from the each of the fibre meshes by aligning the ends of each fibre with the z-axis and connecting the centre of mass of 100 equidistant slices through each fibre, following the approach taken by Lee et al., (2019). This is illustrated in Figure 7.

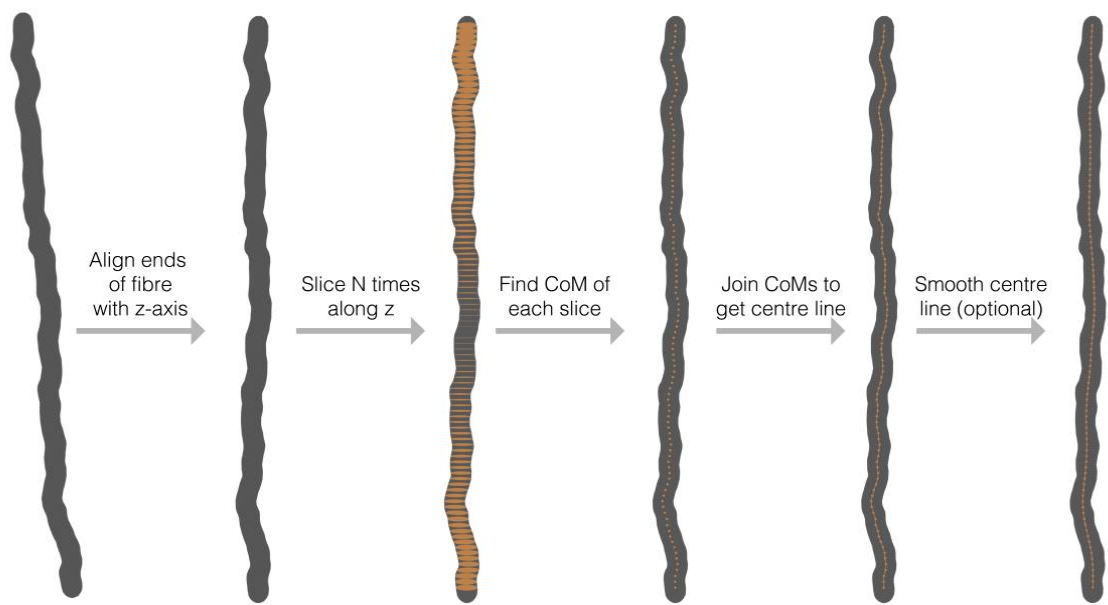

*Figure 7 Centre line extraction of fibres. Each fibre is sliced N times along the z-axis, connecting the centre of mass of each slice to create the points in the centre line. This line can then be optionally smoothed according the diffusion time coarse graining effect, as in (Lee et al., 2019)*

Each segment in this centre line can then be used to assess the microstructure of the phantom. The direction of each segment is used to assess the orientation distribution of the phantom, illustrated in Figure 8. Following the approach of Lee et al., (2019), the direction of each segment is projected onto the surface of a triangulated unit sphere (Womersley, 2018). For each triangle, the number of segments pointing in that direction is used to colour the triangle to visualise the orientation distribution.

A second approach was devised to better visualise orientation distributions in 3D to aid differentiation of crossing bundles and antipodal symmetry. In this approach each vertex is raised above the surface of the sphere proportionally to the number of segments pointing in its direction as illustrated in Figure 8.

To measure the diameter profile along fibres, the direction of each segment gives the normal to a plane used to cut the fibre using Boolean intersection to give a cross section of each fibre at each segment. The diameter profile along the axon is generated by calculating the equivalent diameter of a circle with the same area as the fibre cross section. This process is illustrated in Figure 9.

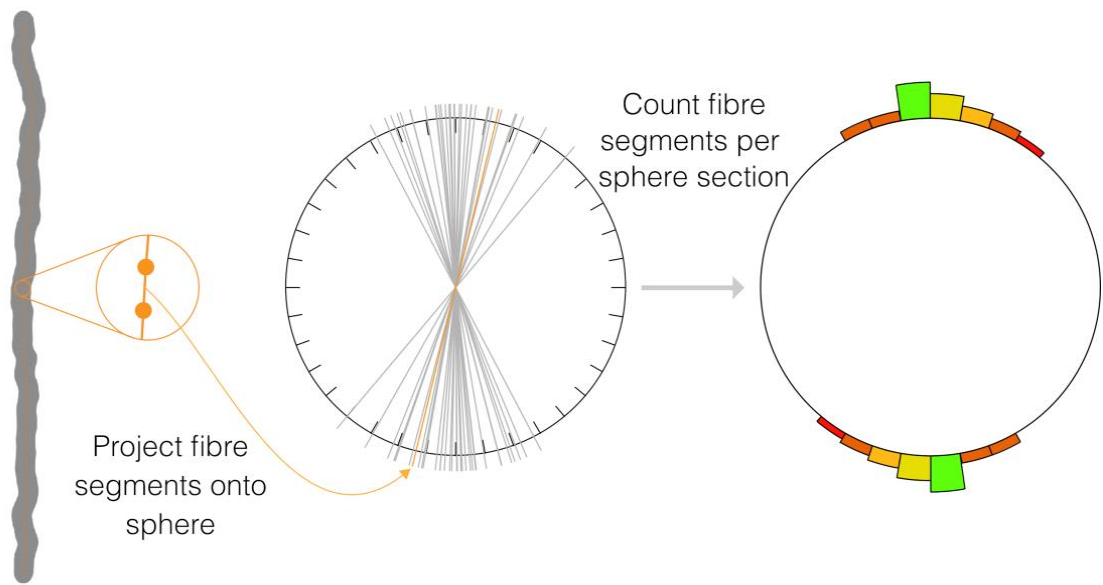

*Figure 8 Orientation distribution calculation. Each segment of a fibre is projected onto the surface of a triangulated sphere, here illustrated with a sectioned circle. For each section in the sphere, the number of fibre segments going through that section Is used to colour and/or raise the surface to visualise the orientation distribution. Since the diffusion process is symmetric about the origin, each fibre segment is projected onto the sphere forwards and backwards.*

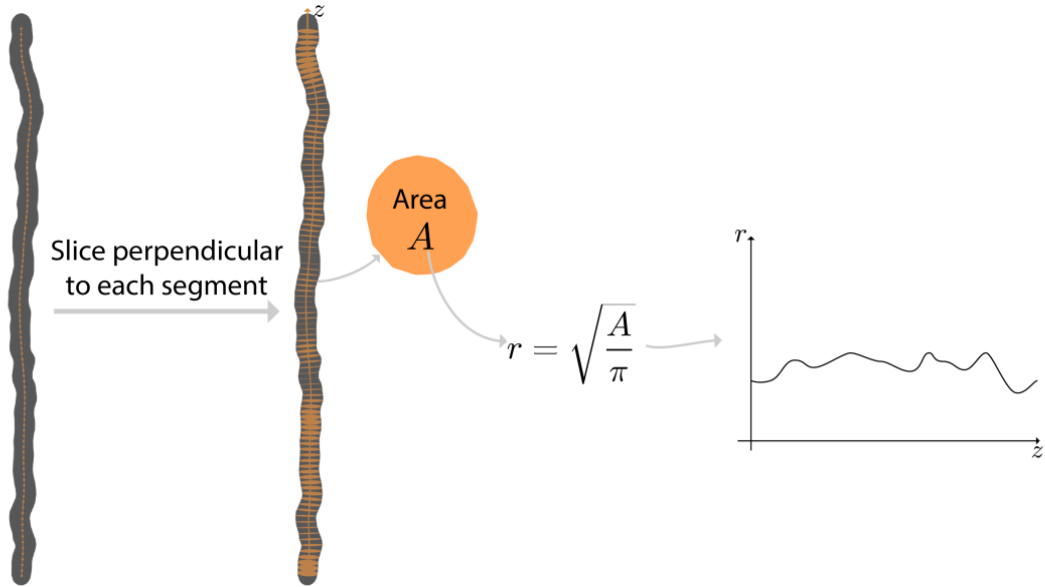

*Figure 9 Calculation of the diameter distribution. A slice is taken through each fibre perpendicular to every segment in the centre line. The area of each of these slices is used to find a circle equivalent radius or diameter using $A = \pi r^2$.*

## 3.3 Virtual Histology

Virtual histological slices are generated to compare ConFiG substrates to real white matter analysed using histology. Histological slices are found by calculating the Boolean intersection of a cutting plane with the generated fibre meshes using Blender (https://blender.org). Virtual histological images are rendered with a resolution of 5nm x 5nm x 100nm, chosen to be comparable to real histological white matter measurements (Abdollahzadeh et al., 2019; Lee et al., 2019).

## 3.4 Relationship between input and output morphology

As mentioned in Section 2, the nature of the ConFiG growth algorithm means that the microstructural morphology of the phantoms may not match the user input. Some fibres may become stuck and fibres cannot typically grow in a straight line, affecting the density and orientation distribution.

To investigate this, we generated a series of ConFiG phantoms with Watson distributed orientation dispersion with $\kappa = [8, 10, 15, 20, 30, 50, 100]$ and target density, $\rho$ = 75%. The mean and standard deviation of the angle from z, $\mu_\theta$ and $\sigma_\theta$ respectively, for each $\kappa$ was calculated by taking 10000 samples from the Watson distribution and this was compared to $\mu_\theta$ and $\sigma_\theta$ of the ConFiG fibres. Additionally, the density of the ConFiG phantoms was compared to the target density of 75%.

### 3.5 Diffusion MRI simulation

To qualitatively verify that the simulated diffusion MRI signals from ConFiG phantoms are realistic, simulated signals from ConFiG phantoms were compared to real HCP data (S. N. Sotiropoulos et al., 2013; Van Essen et al., 2012).

In the real data, the fibre orientation distribution (FOD) was fit in each voxel using constrained spherical deconvolution in MRTrix (Tournier et al., 2019, 2007). Regions of interest (ROIs) were placed in the midbody of the corpus callosum (CC), internal capsule (IC), regions in which a single bundle of fibres is found from the FOD. A third ROI was placed in a region in which three crossing fibre populations were found from visual inspection of the FOD (TC).

In each ROI, the diffusion tensor was fit to the signal and the principle eigenvector used to define a major direction of diffusion in the region, $n$. From this, the normalised diffusion weighted signal was plotted against $|n \cdot G|$, where $G$ is the gradient direction. Additionally, the direction averaged signal was calculated for each b-shell.

To attempt to generate representative microstructure for each ROI using ConFiG, the signal was averaged across each ROI to improve SNR, and the NODDI model (Zhang et al., 2012) was fitted to the average signal to give some initial parameters for ConFiG. Most importantly, the value of $\kappa$ for the Watson distribution (Mardia and Jupp, 2008) estimated using NODDI was used to initialise the orientation dispersion in the ConFiG phantoms used to represent CC ($\kappa = 5.85$) and IC ($\kappa = 4.75$) regions. To represent the TC region, a phantom generated using three mutually perpendicular crossing bundles was used.

ConFiG phantoms were grown using these initial conditions and the diffusion MRI signal simulated using the Camino Monte Carlo diffusion MRI simulator (Hall and Alexander, 2009) with $10^5$ spins and 2000 timesteps. For each phantom, the same processing as with the real data was performed, finding the direction dependent and direction averaged signal per b-shell.

# 4. Results

## 4.1 Impact of biological mechanisms

Each of the proposed biological mechanisms enabled ConFiG to generate phantoms with increased density over the minimal case of chemoattraction only, as is shown in Figure 10. Global optimisation resulted in the in the largest improvement, 17-24%, consistently giving a large improvement. Other improvements performed better for specific phantom configurations. For instance, fasciculation and the dynamic network produced only modest improvements in crossing fibre configurations (4-6%), but performed well in the single bundle cases (11-14%). Fibre collapse was particularly effective in the three perpendicular case, offering 10% improvement.

When combining all of the proposed mechanisms together, the achievable density is higher than any of the improvements individually. This improved performance is comparable to the state of the art, MEDUSA (Ginsburger et al., 2019), with particularly good performance relative to MEDUSA in the crossing fibre configurations.

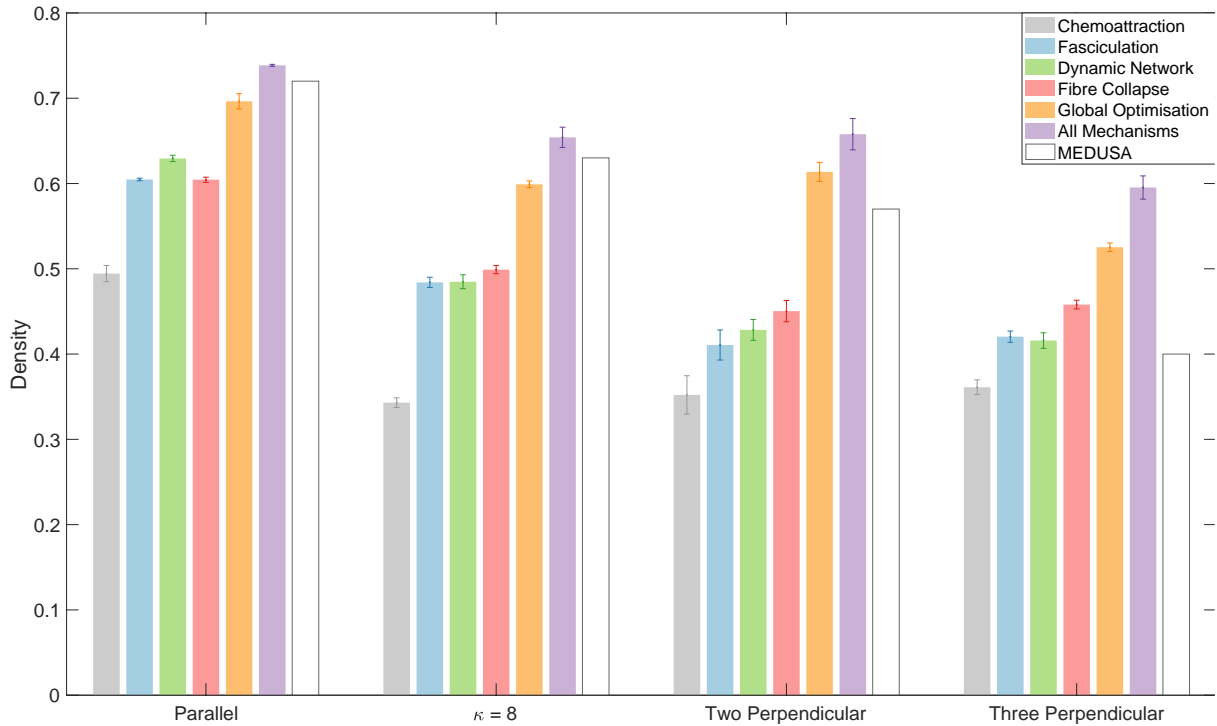

*Figure 10 Demonstration of the impact of each biological growth mechanism on the density achievable with ConFiG. Each bar shows the mean density for each proposed mechanism, error bars show ±standard error on the mean. MEDUSA values are estimated from Fig. 14 in Ginsburger et al.* (Ginsburger et al., 2019).

This improvement in density can be appreciated visually in Figure 11 which demonstrates virtual histology of a parallel fibre phantom for each of the mechanisms. Additionally, Figure 12 visually shows the difference in density of the phantoms in 3D between the minimal case of chemoattraction and all biological mechanism for each fibre configuration.

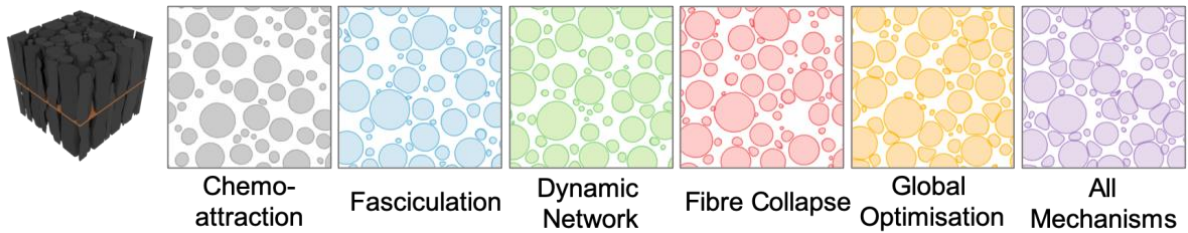

*Figure 11. Virtual histology demonstrating the impact of biologically inspired mechanism on the final phantom created for one of the parallel phantoms tested. This visually demonstrates the improvement in density. Leftmost image shows the phantom generated with all mechanisms in 3D and the cutting plane used to produce the virtual histology.*

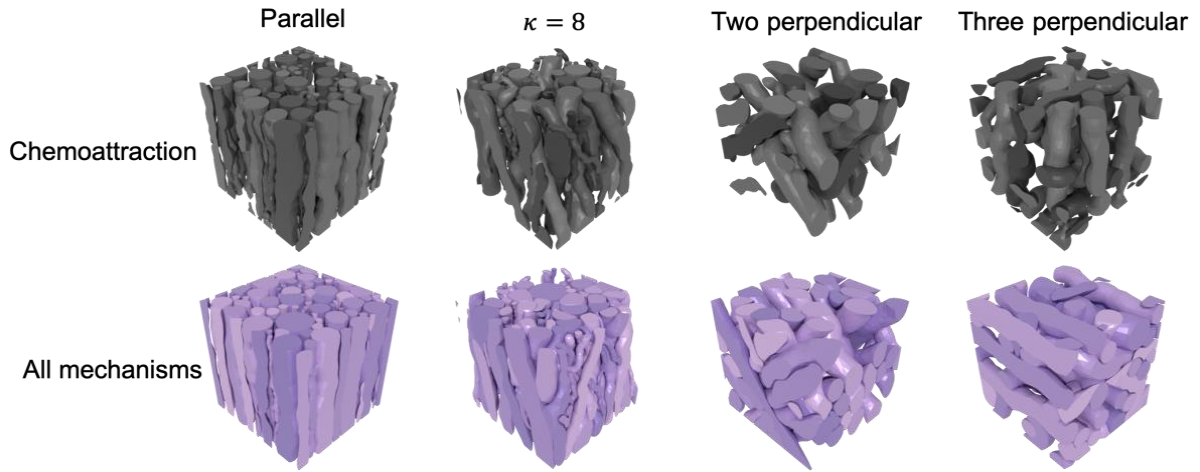

*Figure 12 Demonstration of the improvement in density achieved when using all mechanisms in ConFiG compared to the minimal implementation using only chemoattraction. Colours chosen to match Figure 1.*

The improvement in the density of phantoms leads to a much more realistic simulated diffusion MRI signal as demonstrated in Figure 13. The root mean square error to the real data is reduced by 10 times when using improved ConFiG.

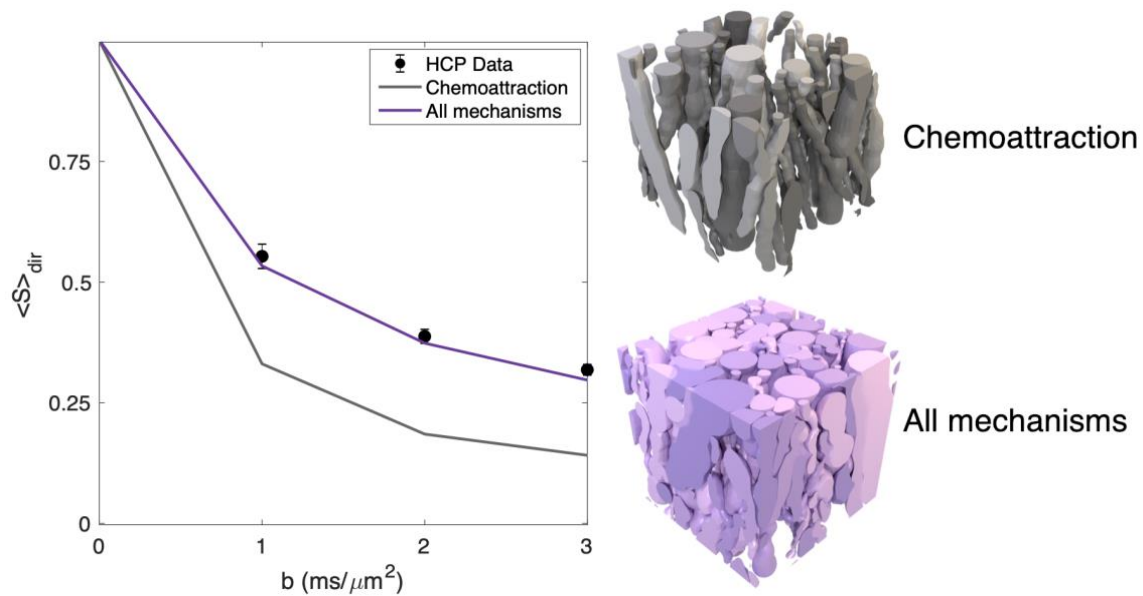

*Figure 13. Left: Direction averaged signal for real HCP data (± standard deviation over ROI) and simulated data from the minimal ConFiG implementation using only chemoattraction and using all growth mechanisms ConFiG showing that ConFiG can produce realistic dMRI signals. Right: The original and improved ConFiG phantoms used to generate the signal on the left. Simulations performed with $10^5$ spins, 2000 timesteps, diffusivity $2.0\mu m^2/ms$ and HCP measurement scheme*(Stamatios N. Sotiropoulos et al., 2013)*.*

## 4.2 Microstructural measures and virtual histology

The microstructural morphology generated using ConFiG is comparable to results from real data as demonstrated in Figures 14-16. Figure 14 demonstrates virtual histology of a ConFiG phantom alongside a real EM image from mouse corpus callosum (Baxi et al., 2015). The exact microstructural features, such as diameter distribution, as well as the EM contrast do not exactly match between ConFiG and the real data. However, ConFiG is able to capture the general morphology or real axons as highlighted in Figure 14. In particular, ConFiG is able to capture complex fibre cross-sections such as in the case of fibres squashed into small spaces. This is the first model of white matter able to handle complex fibre cross-sections such as this to our knowledge.

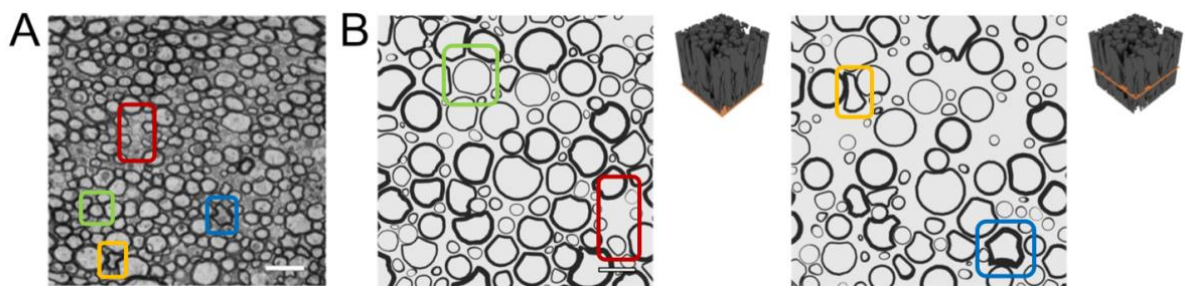

*Figure 14 Comparison of real and virtual histology. A) Light microscopy of rat ventromedial WM in thoracic spinal cord. Reproduced from Baxi et al. 2015* (Baxi et al., 2015)*, scale bar 2μm. B) Two virtual histological slices from a ConFiG generated phantom. Phantoms are rendered to have similar colours to electron microscopy studies. The exact contrast and fibre bundle configurations are different between the real and virtual tissues, but the general morphology of the myelinated axons are captured well using ConFiG as highlighted by corresponding boxes. Yellow and Blue: axons severely deformed between other axons. Red: Pockets of empty space forming. Green: Largely circular axon surrounded by other axons deforming around it. Scale bar 2μm.*

Video 2 show a series of sequential slices through a ConFiG substrate containing two crossing bundles of fibres, demonstrating the non-circular cross-sections generated by ConFiG. These complex cross-sections are not explicitly imposed during growth but arise as a result of the close packing of axons and the meshing procedure used in ConFiG.

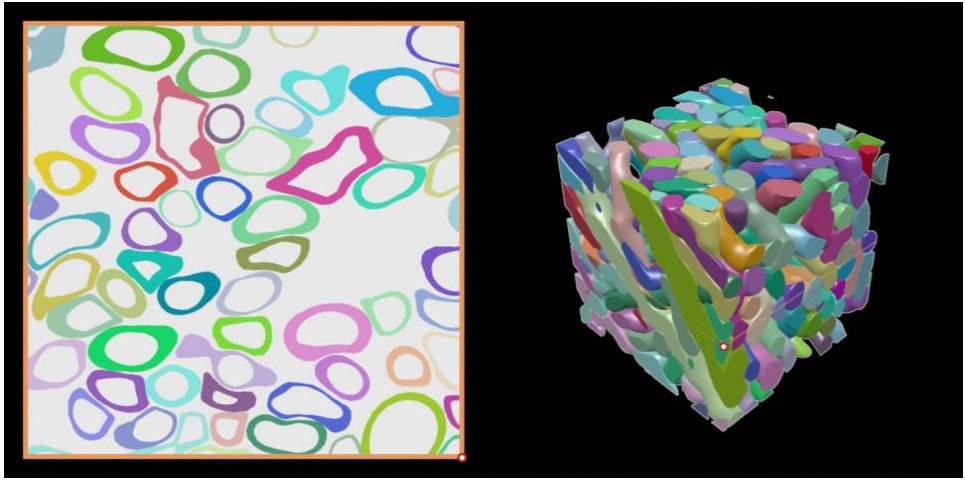

*Video 2 Sequential virtual histology slices of a ConFiG phantom containing two crossing fibre bundles alongside the phantom in 3D with cutting plane indicated. Each fibre is given a unique colour to aid in differentiation and tracking of individual fibre. (Video not available on arxiv, contact corresponding author.)*

The diameter distribution of a ConFiG substrate is compared to a reconstruction from real EM data (Lee et al., 2019) in Figure 15. ConFiG is able to capture the general profile of axonal variations well.

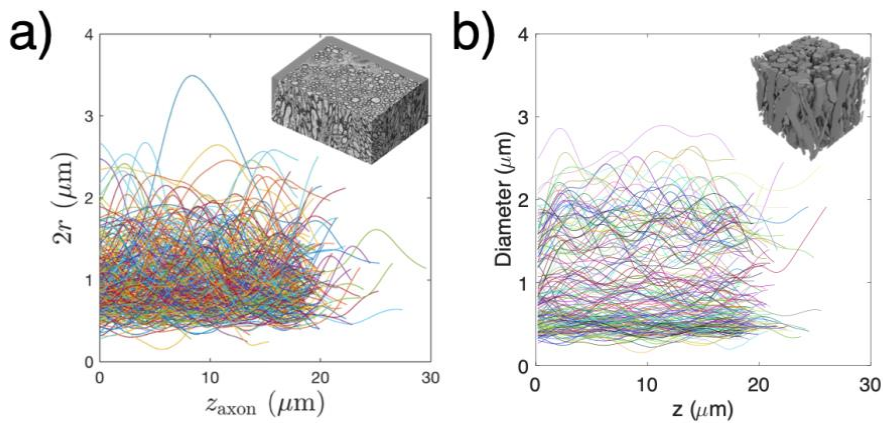

*Figure 15 a) Along fibre diameter variation in ex vivo mouse corpus callosum, reproduced from (Lee et al., 2019). B) Along axis diameter variation in the phantom inset demonstrating the ability of ConFiG to generate realistic microstructure. c) the process used to produce b), the cutting plane is moved along the fibre axis, z, and the area of each slice gives a circle-equivalent diameter.*

ConFiG is also able to generate orientation distributions comparable to real tissue as shown in Figure 16. The orientation dispersion is introduced to ConFiG phantoms using the elliptically symmetrical angular Gaussian (Paine et al., 2018) to best approximate the EM data and also using isotropic Watson distributed directions to demonstrate the flexibility of ConFiG.

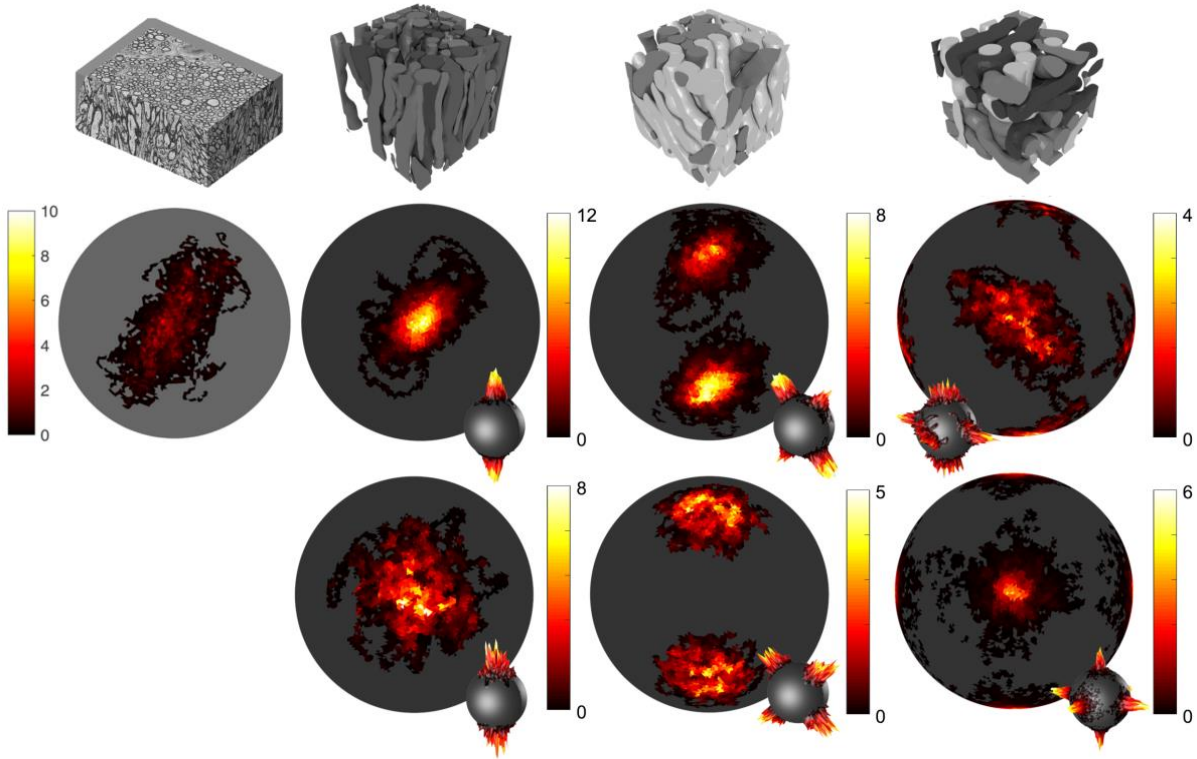

*Figure 16 Dispersion profiles for EM data and a series of numerical phantoms. Top row: EM data used to generate OD profile, reproduced from Lee et al. (Lee et al., 2019) and three ConFiG phantoms with one, two and three crossing bundles (each crossing bundle coloured a different shade of grey). Middle row: OD profile for real EM data and OD profiles corresponding to ConFiG phantoms above, generated using an elliptically symmetric dispersion. Bottom row: Three OD profiles generated from ConFiG phantoms generated using isotropic orientation dispersion. Colormap has units of steradians$^{-1}$.*

## 4.3 Relationship between input and output morphology

*Table 1 Comparison between input microstructural parameters and the microstructure measured in the resulting ConFiG phantoms. For each phantom, an input target density, ρ, of 75% was used with each phantom having a different value of κ*

*used in the Watson distribution. Each κ is associated with a target $\mu_\theta$ and $\sigma_\theta$, the mean and standard deviation of the angle away from the main bundle direction.*

| Input κ | Target $\mu_\theta$ | Output $\mu_\theta$ | Target $\sigma_\theta$ | Output $\sigma_\theta$ | Input ρ | Output ρ |
|---|---|---|---|---|---|---|
| 8 | 19.67 | 17.46 | 11.44 | 9.92 | 75 | 70.6 |
| 10 | 17.12 | 16.47 | 9.67 | 9.43 | 75 | 73.4 |
| 15 | 13.63 | 13.93 | 7.26 | 8.75 | 75 | 73.4 |
| 20 | 11.64 | 12.69 | 6.31 | 7.88 | 75 | 70.7 |
| 30 | 9.41 | 11.60 | 4.97 | 6.60 | 75 | 72.0 |
| 50 | 7.29 | 9.36 | 3.82 | 5.51 | 75 | 73.6 |
| 100 | 5.09 | 7.75 | 2.69 | 4.23 | 75 | 74.9 |

The morphology of ConFiG phantoms matches the input morphology well, as shown in Table 1. Whilst the input and output $\mu_\theta$ and $\sigma_\theta$, do not match exactly, the values are close and increasing the input $\mu_\theta$ and $\sigma_\theta$ also increases the output $\mu_\theta$ and $\sigma_\theta$. Additionally, the output density generally matches the input target density well, achieving higher densities than MEDUSA for the same angular dispersion.

### 4.4 Diffusion MRI simulation

Simulated data from ConFiG substrates match real dMRI data well, as shown in Figure 17. The direction averaged signal matches well in each case, in particular, for the corpus callosum and three crossing phantoms, the simulated signal matches the real signal closely, falling within the error bars (1 standard deviation) of most of the points. The internal capsule is less well represented, with the b=1 ms/μm² signal higher in simulation than in real data.

Additionally, simulated signal as a function of $|n.G|$ the matches the real signal well. In the higher b-value cases, the signal parallel to $n$ reaches the noise floor and the ConFiG simulations begin to deviate from the measured signal.

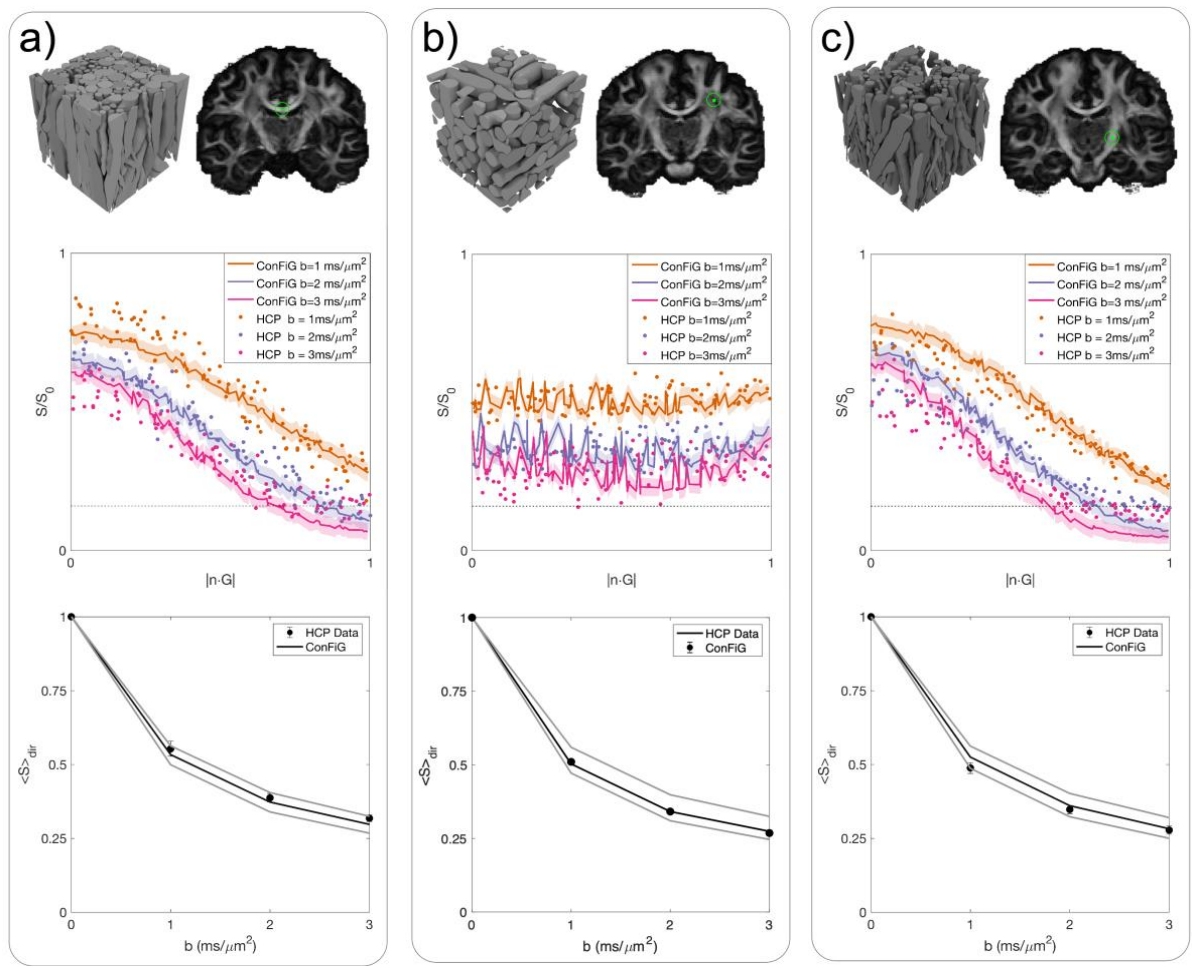

*Figure 17 Comparison of diffusion MRI simulations and real data from three different brain regions: a) an ROI in the midbody of the corpus callosum, with phantom with volume fraction 62% and mean orientation from z 15º. b) an ROI in which there are three crossing bundles, with phantom of three crossing bundles with volume fraction 50% and c) an ROI in the internal capsule, with phantom with volume fraction 60% and mean orientation from z 16.5º. Top row shows the ConFiG phantom and corresponding WM ROI. Middle row shows the direction dependent signal for ConFiG (lines) and HCP data (dots). Bottom row shows the direction averaged signal. Black lines corresponds to phantom in top row. Grey lines are signal from phantoms with the same orientation distribution as the black line in each plot but different densities to show that ConFiG has the flexibility to generate a wide range of realistic signals.*

## 5. Discussion

ConFiG is shown to produce substrates with microstructural properties comparable to real white matter, both in terms of measures derived from histology (i.e., electron microscopy) and in terms of the diffusion MRI signal.

ConFiG is shown to produce WM numerical phantoms with state-of-the-art performance. The amount of real data containing 3D microstructural morphology information available to compare to is limited, so we have only compared to one sample in this study. Whilst limited, this shows that ConFiG is able to produce realistic microstructure by following simple biologically inspired growth rules.

Whilst the input morphological priors do not necessarily correspond to the morphology of the resulting ConFiG phantom, Table 1 shows that even for relatively high orientation dispersion and density, this effect is small. Even so, for use in further analysis, microstructural measures such as orientation dispersion and density should be calculated based on the resultant phantom, rather than taking the input microstructural parameters.

A related property of ConFiG is that the growth algorithm strongly depends on the growth network, meaning that the resulting phantom for the same input fibre configuration will be different for different network choices. This is alleviated to an extent by using the dynamic network introduced here, however the phantom will still be dependent on the initialisation of the network. The dependence appears to be relatively minor as is demonstrated by the small standard errors on the mean density shown in Figure 10 across the five repetitions.

The diffusion MRI simulations shown in Figure 17 demonstrate the ability of ConFiG to generate phantoms which reproduce real diffusion MRI data well. These simulations, however, are just three examples of ConFiG phantoms and corresponding simulations. Using NODDI as input to ConFiG means that the resulting phantoms have sensible morphologies and are shown to generate signals that match the real tissue well, though there may be other configurations that can better reproduce the signal. It may be possible to find a better matching phantom using a computational modelling approach such as that proposed in (Nedjati-Gilani et al., 2017), however the simulations presented are sufficient to demonstrate a proof-of-concept that ConFiG can be used to generate realistic simulated dMRI data.

**Limitations and Future Work**

One limitation of ConFiG is that the algorithm relies on the space being sufficiently densely sampled by the growth network. This can require a large number of nodes for a large phantom, becoming prohibitively memory expensive. The dependence of the resulting phantom on the density of network nodes can be addressed by growing the fibres in small subregions local to the head of the fibres rather than the whole space at once. For instance, rather than filling the entire space of growth with nodes, it is possible to fill a small layer of the space with points and then grow layer by layer. In this way, it is possible to achieve a high density of nodes using fewer nodes than when covering the entire space.

One further potential limitation of ConFiG is that once a fibre has grown, it is static. The fibre will remain fixed in place and all other fibres will have to grow around it. One problem with this is that once the fibres are fixed, they may create pockets of inaccessible space which limits the space available for following fibres. Additionally, in real tissue, axons are flexible and non-rigid, meaning that it may be more realistic that growing fibres can push existing fibres out of the way to make more space for growth. A potential approach to ameliorate this would be to have an optimisation procedure during growth, similar to the global optimisation introduced in this work but optimising the shape of a fibre as it grows.

A limitation of the current study is that the simulations assume a single diffusivity for the intra and extracellular spaces and no permeability of the axonal membranes. Furthermore, effects such as T2 and magnetic susceptibility are ignored. These effects are a limitation of the simulator used rather than ConFiG, and work is planned to improve these aspects of the simulator for more realistic simulated signals.

Additionally, as mentioned above, this study only compares ConFiG to one EM sample of real tissue. Future work will also aim at more extensive validation of the digital phantoms generated using ConFiG, making comparison with larger EM dataset, including different WM configurations from different brain regions.

We will work towards decreasing the difference between the input and output morphological measures, particularly in complex situations, such as high orientation dispersion and crossing bundles. This can be addressed through the improvements to ConFiG mentioned here and also by improving the strategy for the generation of starting and target points for each fibre. For instance, currently it is not intuitive how starting and target points should be arranged to achieve a desired density in crossing regions of fibres.

One planned extension of ConFiG is to implement periodic boundary conditions in the growth network, enabling the generation of fully periodic phantoms. This would enable ConFiG phantoms to be generated in relatively small volumes and tiled for simulation, accelerating the process of generating a wide range of phantoms and the memory required to store each phantom.

The core growth algorithm for ConFiG relies on a set of starting and target points, a connected network of nodes and some rules defining the growth. As such, ConFiG is very flexible since the exact form of each of these components can be modified based on the application. One example of a simple modification that may be explored is the order of growth of the axons. Currently, in the absence of any clear biological precedent know to the authors, fibres grow in a random order, but it may be possible that there is a better order such as growing large diameter axons first, or central axons in a bundle growing first.

In this work, ConFiG is applied to the case of densely packed axons, without contributions from neuronal cell bodies or other processes. A planned future extension of ConFiG is to allow for the addition of glial cells such as astrocytes and oligodendrocytes (e.g. from real 3D reconstructions available for instance on http://neuromorpho.org or synthetically generated using generative models like in (Palombo et al., 2019)) to the extracellular space to make the virtual WM tissue more realistic.

Additionally, to further add to the realism of ConFiG phantoms, realistic myelin may be modelled, creating spiral layers wrapped around the axons (Brusini et al., 2019). Furthermore, intra-axonal structures such as mitochondria and microtubules may be added to investigate their contributions to the diffusion weighted signal.

A planned future application will be to use ConFiG to generate a wide range of phantoms with different microstructural features. These can then be used to create a computational model to estimate microstructural features from the diffusion MRI signal in an approach similar to previous works (Hill et al., 2019; Nedjati-Gilani et al., 2017; Palombo et al., 2018a; Rensonnet et al., 2018).

**Applications beyond diffusion MRI**

As mentioned in the introduction, axonal configuration impacts MR signals beyond dMRI. One potential avenue of exploration would be to investigate the impact of realistic axonal configurations in a similar way to Xu et al. (Xu et al., 2018), extending their 2D simulations to use realistic 3D geometries generated in ConFiG.

The virtual histology presented in Figure 11 show an approximation of electron microscopy generated using ConFiG substrates. In this work, the purpose of this is to show that ConFiG is generating microstructurally realistic phantoms. For this reason, the virtual histology is simply produced by rendering images to have similar contrast to electron microscopy for comparison. It may be possible, however to generate more realistic electron microscopy images using a physically realistic electron microscopy simulator (Babin et al., 2010; Grella et al., 2003; Ophus, 2017) which may be used to train and test axon segmentation routines. This may be of particular use for cases of fibres parallel to the electron microscopy plane or crossing bundles which are typically difficult for 3D reconstruction and segmentation algorithms.

The 3D meshes generated by ConFiG are saved in the PLY format, a widely used format for storing meshes for many purposes. This means that the ConFiG phantoms may be used in other types of simulations such as polarized light imaging (Matuschke, 2019; Menzel et al., 2015) or molecular dynamics simulations using software such as MCell (https://mcell.org) (Kerr et al., 2008; Stiles et al., 1996; Stiles and Bartol, 2001) or LAMMPS (http://lammps.sandia.gov) (Plimpton, 1997).

# 6. Conclusion

ConFiG enables the generation of realistic white matter numerical phantoms achieving state of the art fibre density whilst ensuring realistic microstructural morphology by following biologically motivated rules. This realistic microstructure is shown to generate realistic simulated diffusion MRI signals, opening up the possibility to use ConFiG to create a realistic computational model of WM microstructure.

ConFiG outputs fibre meshes which can be used for realistic diffusion MRI simulations or can be processed to produce virtual histological slices, allowing for further potential applications outside of diffusion MRI.

# Author Credit

**Ross Callaghan:** Conceptualisation, Methodology, Software, Investigation, Visualisation, Writing – Original Draft. **Daniel C. Alexander:** Conceptualisation, Writing – Review & Editing, Supervision, Resources, Funding acquisition. **Marco Palombo:** Conceptualisation, Methodology, Writing – Review & Editing, Supervision. **Hui Zhang:** Conceptualisation, Methodology, Writing – Review & Editing, Supervision, Resources, Funding acquisition.

# Data and Code Availability

ConFiG code will be made available at https://github.com/ucl-mig.

# Conflicts of Interest

The authors confirm that there are no conflicts commercial or financial conflicts of interest affecting this work.

# Acknowledgements

This work is supported by the EPSRC-funded UCL Centre for Doctoral Training in Medical Imaging (EP/L016478/1) and the Department of Health's NIHR-funded Biomedical Research Centre at University College London Hospitals. This work was supported by EPSRC grants EP/M020533/1 and EP/N018702/1.

Data were provided by the Human Connectome Project, WU-Minn Consortium (Principal Investigators: David Van Essen and Kamil Ugurbil; 1U54MH091657) funded by the 16 NIH

Institutes and Centers that support the NIH Blueprint for Neuroscience Research; and by the McDonnell Center for Systems Neuroscience at Washington University.**REFERENCES**

Abdollahzadeh, A., Ilya, B., Jokitalo, E., Tohka, J., Sierra, A., 2019. Automated 3D Axonal Morphometry of White Matter. Sci. Rep. 9, 1–16. https://doi.org/10.1038/s41598-019-42648-2

Alexander, D.C., Hubbard, P.L., Hall, M.G., Moore, E.A., Ptito, M., Parker, G.J.M., Dyrby, T.B., 2010. Orientationally invariant indices of axon diameter and density from diffusion MRI. Neuroimage 52, 1374–1389. https://doi.org/10.1016/j.neuroimage.2010.05.043

Babin, S., Borisov, S.S., Ito, H., Ivanchikov, A., Suzuki, M., 2010. Simulation of scanning electron microscope images taking into account local and global electromagnetic fields. J. Vac. Sci. Technol. B, Nanotechnol. Microelectron. Mater. Process. Meas. Phenom. 28, C6C41-C6C47. https://doi.org/10.1116/1.3518917

Barry, J., Gu, Y., Gu, C., 2010. Polarized targeting of L1-CAM regulates axonal and dendritic bundling in vitro. Eur. J. Neurosci. 32, 1618–1631. https://doi.org/10.1111/j.1460-9568.2010.07447.x

Baxi, E.G., DeBruin, J., Tosi, D.M., Grishkan, I. V., Smith, M.D., Kirby, L.A., Strasburger, H.J., Fairchild, A.N., Calabresi, P.A., Gocke, A.R., 2015. Transfer of myelin-reactive Th17 cells impairs endogenous remyelination in the central nervous system of cuprizone-fed mice. J. Neurosci. 35, 8626–8639. https://doi.org/10.1523/JNEUROSCI.3817-14.2015

Brabec, J., Lasič, S., Nilsson, M., 2019. Time-dependent diffusion in undulating thin fibers: Impact on axon diameter estimation. NMR Biomed. 1–19. https://doi.org/10.1002/nbm.4187

Brusini, L., Menegaz, G., Nilsson, M., 2019. Monte Carlo simulations of water exchange through myelin wraps: Implications for diffusion MRI. IEEE Trans. Med. Imaging 38, 1438–1445. https://doi.org/10.1109/TMI.2019.2894398

Budde, M.D., Frank, J.A., 2010. Neurite beading is sufficient to decrease the apparent diffusion coefficient after ischemic stroke. PNAS 107, 14472–14477. https://doi.org/10.1073/pnas.1004841107/-/DCSupplemental.www.pnas.org/cgi/doi/10.1073/pnas.1004841107

Callaghan, R., Alexander, D.C., Zhang, H., Palombo, M., 2019. Contextual Fibre Growth to Generate Realistic Axonal Packing for Diffusion MRI Simulation, in: Chung, A., Gee, J., Yushkevich, P., Bao, S. (Eds.), Information Processing in Medical Imaging. IPMI2019. Lecture Notes in Computer Science, Vol 11492. Springer, Cham. https://doi.org/10.1007/978-3-030-20351-1_33

Close, T.G., Tournier, J.D., Calamante, F., Johnston, L.A., Mareels, I., Connelly, A., 2009. A software tool to generate simulated white matter structures for the assessment of fibre-tracking algorithms. Neuroimage 47, 1288–1300. https://doi.org/10.1016/j.neuroimage.2009.03.077

Cook, P. a, Bai, Y., Seunarine, K.K., Hall, M.G., Parker, G.J., Alexander, D.C., 2006. Camino: Open-Source Diffusion-MRI Reconstruction and Processing. 14th Sci. Meet. Int. Soc. Magn. Reson. Med. 14, 2759.

Dent, E.W., Gupton, S.L., Gertler, F.B., 2011. The growth cone cytoskeleton in Axon outgrowth and guidance. Cold Spring Harb. Perspect. Biol. 3, 1–39.36

https://doi.org/10.1101/cshperspect.a001800

Fieremans, E., Novikov, D.S., Jensen, J.H., Helpern, J.A., 2010. Monte Carlo study of a two-compartment exchange model of diffusion. NMR Biomed. 23, 711–724. https://doi.org/10.1002/nbm.1577

Ford, J.C., Hackney, D.B., 1997. Numerical model for calculation of apparent diffusion coefficients (ADC) in permeable cylinders - Comparison with measured ADC in spinal cord white matter. Magn. Reson. Med. 37, 387–394. https://doi.org/10.1002/mrm.1910370315

Ginsburger, K., Matuschke, F., Poupon, F., Mangin, J.-F., Axer, M., Poupon, C., 2019. MEDUSA : A GPU-based tool to create realistic phantoms of the brain microstructure using tiny spheres. Neuroimage 193, 10–24. https://doi.org/10.1016/j.neuroimage.2019.02.055

Ginsburger, K., Poupon, F., Beaujoin, J., Estournet, D., Matuschke, F., Mangin, J.-F., Axer, M., Poupon, C., 2018. Improving the realism of white matter numerical phantoms: A step toward a better understanding of the influence of structural disorders in diffusion MRI. Front. Phys. 5, 1–18. https://doi.org/10.3389/fphy.2018.00012

Grella, L., Lorusso, G., Adler, D.L., 2003. Simulations of Scanning Electron Microscopy Imaging and Charging of Insulating Structures. Scanning 25, 300–308.

Hall, M.G., Alexander, D.C., 2009. Convergence and Parameter Choice for Monte-Carlo Simulations of Diffusion MRI. IEEE Trans. Med. Imaging 28, 1354–1364. https://doi.org/10.1109/TMI.2009.2015756

Hill, I., Palombo, M., Santin, M., Branzoli, F., Philippe, A., 2019. Machine learning based white matter models with permeability : An experimental study in cuprizone treated in-vivo mouse model of axonal demyelination. ArXiv.

Jelescu, I.O., Budde, M.D., 2017. Design and Validation of Diffusion MRI Models of White Matter. Front. Phys. 5. https://doi.org/10.3389/fphy.2017.00061

Kerr, R.A., Bartol, T.M., Kaminsky, B., Dittrich, M., Chang, J.-C.J., Baden, S.B., Sejnowski, T.J., Stiles, J.R., 2008. Fast Monte Carlo Simulation Methods for Biological Reaction-Diffusion Systems in Solution and on Surfaces. SIAM J Sci Comput 30, 3126. https://doi.org/http://dx.doi.org/10.1137/070692017

Landman, B.A., Farrell, J.A.D., Smith, S.A., Reich, D.S., Calabresi, P.A., Van Zijl, P.C.M., 2010. Complex geometric models of diffusion and relaxation in healthy and damaged white matter. NMR Biomed. 23, 152–162. https://doi.org/10.1002/nbm.1437

Lee, H.H., Yaros, K., Veraart, J., Pathan, J.L., Xia, F., Sungheon, L., Novikov, D.S., Fieremans, E., 2019. Along-axon diameter variation and axonal orientation dispersion revealed with 3D electron microscopy : implications for quantifying brain white matter microstructure with histology and diffusion MRI. Brain Struct. Funct. 0, 0. https://doi.org/10.1007/s00429-019-01844-6

Lee, J., Shmueli, K., Fukunaga, M., Van Gelderen, P., Merkle, H., Silva, A.C., Duyn, J.H., 2010. Sensitivity of MRI resonance frequency to the orientation of brain tissue microstructure. Proc. Natl. Acad. Sci. U. S. A. 107, 5130–5135. https://doi.org/10.1073/pnas.0910222107

Li, J.R., Nguyen, V.D., Tran, T.N., Valdman, J., Trang, C.B., Nguyen, K. Van, Vu, D.T.S., Tran, H.A., Tran, H.T.A., Nguyen, T.M.P., 2019. SpinDoctor: A MATLAB toolbox for diffusion MRI simulation. Neuroimage 202, 116120. https://doi.org/10.1016/j.neuroimage.2019.116120

Li, W., Wu, B., Avram, A. V., Liu, C., 2012. Magnetic susceptibility anisotropy of human brain in vivo and its molecular underpinnings. Neuroimage 59, 2088–2097. https://doi.org/10.1016/j.neuroimage.2011.10.038

Lowery, L.A., Vactor, D. Van, 2009. The trip of the tip: Understanding the growth cone machinery. Nat. Rev. Mol. Cell Biol. 10, 332–343. https://doi.org/10.1038/nrm2679

Mardia, K. V., Jupp, P.E., 2008. Directional Statistics, Directional Statistics. John Wiley & Sons, Ltd. https://doi.org/10.1002/9780470316979

Matuschke, F., 2019. Dense Fiber Modeling for 3D-Polarized Light Imaging Simulations. ArXiv.

Menzel, M., Michielsen, K., De Raedt, H., Reckfort, J., Amunts, K., Axer, M., 2015. A Jones matrix formalism for simulating three-dimensional polarized light imaging of brain tissue. J. R. Soc. Interface 12. https://doi.org/10.1098/rsif.2015.0734

Mortimer, D., Fothergill, T., Pujic, Z., Richards, L.J., Goodhill, G.J., 2008. Growth cone chemotaxis. Trends Neurosci. 31, 90–98. https://doi.org/10.1016/j.tins.2007.11.008

Nedjati-Gilani, G.L., Schneider, T., Hall, M.G., Cawley, N., Hill, I., Ciccarelli, O., Drobnjak, I., Wheeler-Kingshott, C.A.M.G., Alexander, D.C., 2017. Machine learning based compartment models with permeability for white matter microstructure imaging. Neuroimage. https://doi.org/10.1016/j.neuroimage.2017.02.013

Neher, P.F., Laun, F.B., Stieltjes, B., Maier-Hein, K.H., 2014. Fiberfox: Facilitating the creation of realistic white matter software phantoms. Magn. Reson. Med. 72, 1460–1470. https://doi.org/10.1002/mrm.25045

Nilsson, M., Alerstam, E., Wirestam, R., Ståhlberg, F., Brockstedt, S., Lätt, J., 2010. Evaluating the accuracy and precision of a two-compartment Karger model using Monte Carlo simulations. J. Magn. Reson. 206, 59–67. https://doi.org/10.1016/j.jmr.2010.06.002

Nilsson, M., Lasič, S., Drobnjak, I., Topgaard, D., Westin, C.F., 2017. Resolution limit of cylinder diameter estimation by diffusion MRI: The impact of gradient waveform and orientation dispersion. NMR Biomed. 30, 1–13. https://doi.org/10.1002/nbm.3711

Nilsson, M., Lätt, J., Nordh, E., Wirestam, R., Ståhlberg, F., Brockstedt, S., 2009. On the effects of a varied diffusion time in vivo: is the diffusion in white matter restricted? Magn. Reson. Imaging 27, 176–187. https://doi.org/10.1016/j.mri.2008.06.003

Nilsson, M., Lätt, J., Ståhlberg, F., van Westen, D., Hagslätt, H., 2012. The importance of axonal undulation in diffusion MR measurements: A Monte Carlo simulation study. NMR Biomed. 25, 795–805. https://doi.org/10.1002/nbm.1795

Ophus, C., 2017. A fast image simulation algorithm for scanning transmission electron microscopy. Adv. Struct. Chem. Imaging 3, 1–11. https://doi.org/10.1186/s40679-017-0046-1

Paine, P.J., Preston, S.P., Tsagris, M., Wood, A.T.A., 2018. An elliptically symmetric angular Gaussian distribution. Stat. Comput. 28, 689–697. https://doi.org/10.1007/s11222-017-9756-4

Palombo, M., Alexander, D.C., Zhang, H., 2019. A generative model of realistic brain cells with application to numerical simulation of diffusion- weighted MR signal. Neuroimage 188, 391–402. https://doi.org/S1053811918321694

Palombo, M., Hill, I., Santin, M.D., Branzoli, F., Philippe, A.-C., Wassermann, D., Aigrot, M.-S., Stankoff, B., Zhang, H., Lehericy, S., Petiet, A., Alexander, D.C., Drobnjak, I., 2018a. Machine learning based estimation of axonal permeability: validation on cuprizone treated in-vivo mouse model of axonal demyelination, in: Proc. Joint Annual Meeting ISMRM-ESMRMB, Paris, France.

Palombo, M., Ligneul, C., Hernandez-Garzon, E., Valette, J., 2018b. Can we detect the effect

of spines and leaflets on the diffusion of brain intracellular metabolites? Neuroimage. https://doi.org/10.1016/j.neuroimage.2017.05.003

Plimpton, S., 1997. Short-Range Molecular Dynamics. J. Comput. Phys. 117, 1–42. https://doi.org/10.1006/jcph.1995.1039

Polleux, F., Snider, W., 2010. Initiating and growing an axon. Cold Spring Harb. Perspect. Biol. 2, 1–20. https://doi.org/10.1101/cshperspect.a001925

Price, D.J., Jarman, A.P., Mason, J.O., Kind, P.C., 2017. Building brains - An introduction to neural development, Building Brains - An Introduction to Neural Development. https://doi.org/10.1002/9781119293897

Rafael-Patino, J., Girard, G., Romascano, D., Barakovic, M., Rensonnet, G., Thiran, J.-P., Daducci, A., 2018. Realistic 3D Fiber Crossing Phantom Models for Monte Carlo Diffusion Simulations, in: 26th Annual Meeting of the International Society for Magnetic Resonance in Medicine (ISMRM).

Rauch, P., Heine, P., Goettgens, B., Käs, J.A., 2013. Different modes of growth cone collapse in NG 108-15 cells. Eur. Biophys. J. 42, 591–605. https://doi.org/10.1007/s00249-013-0907-z

Rensonnet, G., Scherrer, B., Girard, G., Jankovski, A., Warfield, S.K., Macq, B., Thiran, J.-P., Taquet, M., 2018. Towards microstructure fingerprinting: Estimation of tissue properties from a dictionary of Monte Carlo diffusion MRI simulations. Neuroimage 184, 964–980. https://doi.org/10.1016/J.NEUROIMAGE.2018.09.076

Rensonnet, G., Scherrer, B., Warfield, S.K., Macq, B., Taquet, M., 2017. Assessing the validity of the approximation of diffusion-weighted-MRI signals from crossing fascicles by sums of signals from single fascicles. Magn. Reson. Med. 2345, 2332–2345. https://doi.org/10.1002/mrm.26832

Sakisaka, T., Takai, Y., 2005. Cell adhesion molecules in the CNS. J. Cell Sci. 118, 5407–5410. https://doi.org/10.1242/jcs.02672

Salo, R.A., Belevich, I., Manninen, E., Jokitalo, E., Gröhn, O., Sierra, A., 2018. Quantification of anisotropy and orientation in 3D electron microscopy and diffusion tensor imaging in injured rat brain. Neuroimage 172, 404–414. https://doi.org/10.1016/j.neuroimage.2018.01.087

Scherrer, B., Schwartzman, A., Taquet, M., Sahin, M., Prabhu, S.P., Warfield, S.K., 2016. Characterizing brain tissue by assessment of the distribution of anisotropic microstructural environments in diffusion-compartment imaging (DIAMOND). Magn. Reson. Med. 76, 963–977. https://doi.org/10.1002/mrm.25912

Šmít, D., Fouquet, C., Pincet, F., Zapotocky, M., Trembleau, A., 2017. Axon tension regulates fasciculation/defasciculation through the control of axon shaft zippering. Elife 6, 1–49. https://doi.org/10.7554/eLife.19907

Sotiropoulos, Stamatios N., Jbabdi, S., Xu, J., Andersson, J.L., Moeller, S., Auerbach, E.J., Glasser, M.F., Hernandez, M., Sapiro, G., Jenkinson, M., Feinberg, D.A., Yacoub, E., Lenglet, C., Van Essen, D.C., Ugurbil, K., Behrens, T.E.J., 2013. Advances in diffusion MRI acquisition and processing in the Human Connectome Project. Neuroimage 80, 125–143. https://doi.org/10.1016/j.neuroimage.2013.05.057

Sotiropoulos, S. N., Moeller, S., Jbabdi, S., Xu, J., Andersson, J.L., Auerbach, E.J., Yacoub, E., Feinberg, D., Setsompop, K., Wald, L.L., Behrens, T.E.J., Ugurbil, K., Lenglet, C., 2013. Effects of image reconstruction on fiber orientation mapping from multichannel diffusion MRI: Reducing the noise floor using SENSE. Magn. Reson. Med. 70, 1682–1689. https://doi.org/10.1002/mrm.24623

Stiles, J.R., Bartol, T.M., 2001. Monte Carlo Methods for Simulating Realistic Synaptic Microphysiology using MCELL, Computational Neuroscience: Realistic Modeling for Experimentalists. https://doi.org/10.1201/9781420039290.ch4

Stiles, J.R., Van Helden, D., Bartol, T.M., Salpeter, E.E., Salpeter, M.M., 1996. Miniature endplate current rise times less than 100 microseconds from improved dual recordings can be modeled with passive acetylcholine diffusion from a synaptic vesicle. Proc. Natl. Acad. Sci. 93, 5747–5752. https://doi.org/10.1073/pnas.93.12.5747

Tariq, M., Schneider, T., Alexander, D.C., Gandini Wheeler-Kingshott, C.A., Zhang, H., 2016. Bingham-NODDI: Mapping anisotropic orientation dispersion of neurites using diffusion MRI. Neuroimage 133, 207–223. https://doi.org/10.1016/j.neuroimage.2016.01.046

Tournier, J.D., Calamante, F., Connelly, A., 2007. Robust determination of the fibre orientation distribution in diffusion MRI: Non-negativity constrained super-resolved spherical deconvolution. Neuroimage 35, 1459–1472. https://doi.org/10.1016/j.neuroimage.2007.02.016

Tournier, J.D., Smith, R., Raffelt, D., Tabbara, R., Dhollander, T., Pietsch, M., Christiaens, D., Jeurissen, B., Yeh, C.H., Connelly, A., 2019. MRtrix3: A fast, flexible and open software framework for medical image processing and visualisation. Neuroimage 202, 116137. https://doi.org/10.1016/j.neuroimage.2019.116137

Van Essen, D.C., Ugurbil, K., Auerbach, E., Barch, D., Behrens, T.E.J., Bucholz, R., Chang, A., Chen, L., Corbetta, M., Curtiss, S.W., Della Penna, S., Feinberg, D., Glasser, M.F., Harel, N., Heath, A.C., Larson-Prior, L., Marcus, D., Michalareas, G., Moeller, S., Oostenveld, R., Petersen, S.E., Prior, F., Schlaggar, B.L., Smith, S.M., Snyder, A.Z., Xu, J., Yacoub, E., 2012. The Human Connectome Project: A data acquisition perspective. Neuroimage. https://doi.org/10.1016/j.neuroimage.2012.02.018

Voyiadjis, A.G., Doumi, M., Curcio, E., Shinbrot, T., 2011. Fasciculation and defasciculation of neurite bundles on micropatterned substrates. Ann. Biomed. Eng. 39, 559–569. https://doi.org/10.1007/s10439-010-0168-2

Womersley, R.S., 2018. Efficient spherical designs with good geometric properties. Contemp. Comput. Math. - A Celebr. 80th Birthd. Ian Sloan 1243–1285. https://doi.org/10.1007/978-3-319-72456-0_57

Xu, J., Li, H., Harkins, K.D., Jiang, X., Xie, J., Kang, H., Does, M.D., Gore, J.C., 2014. Mapping mean axon diameter and axonal volume fraction by MRI using temporal diffusion spectroscopy. Neuroimage 103, 10–19. https://doi.org/10.1016/j.neuroimage.2014.09.006

Xu, T., Foxley, S., Kleinnijenhuis, M., Chen, W.C., Miller, K.L., 2018. The effect of realistic geometries on the susceptibility-weighted MR signal in white matter. Magn. Reson. Med. 79, 489–500. https://doi.org/10.1002/mrm.26689

Zhang, H., Hubbard, P.L., Parker, G.J.M., Alexander, D.C., 2011. Axon diameter mapping in the presence of orientation dispersion with diffusion MRI. Neuroimage 56, 1301–1315. https://doi.org/10.1016/j.neuroimage.2011.01.084

Zhang, H., Schneider, T., Wheeler-Kingshott, C.A., Alexander, D.C., 2012. NODDI: Practical in vivo neurite orientation dispersion and density imaging of the human brain. Neuroimage 61, 1000–1016. https://doi.org/10.1016/j.neuroimage.2012.03.072